\documentclass[preprint,preprintnumbers,amsmath,amssymb]{revtex4-1}

\usepackage{graphicx,tabularx}
\usepackage{bm}
\usepackage{float}
\usepackage{dcolumn}
\usepackage{color}
\usepackage{ulem}
\usepackage{mathptmx}
\usepackage{moreverb}
\usepackage[symbol]{footmisc}

\definecolor{darkpastelgreen}{rgb}{0.01, 0.75, 0.24}

\renewcommand{\thefootnote}{\fnsymbol{footnote}}


\begin{document}
\renewcommand{\figurename}{\textbf{Fig.}}
\renewcommand{\thefigure}{\textbf{\arabic{figure}}}
\title{Unconventional Magnetic Oscillations in Kagome Mott Insulators}

\date{\today}


\author{Guoxin Zheng$^{1\dagger}$, Yuan Zhu$^{1\dagger}$, Kuan-Wen Chen$^{1\dagger}$, Byungmin Kang$^2$, Dechen Zhang$^{1}$, Kaila Jenkins$^{1}$, Aaron Chan$^{1}$, Zhenyuan Zeng$^{3,4}$, Aini Xu$^{3,4}$, Oscar A. Valenzuela$^{5}$, Joanna Blawat$^{5}$, John Singleton$^{5}$, Patrick A. Lee$^2$, Shiliang Li$^{3,4,6}$}
\author{Lu Li$^{1}$}
\email{luli@umich.edu}

\affiliation{
	$^1$Department of Physics, University of Michigan, Ann Arbor, MI 48109, USA\\
	$^2$Department of Physics, Massachusetts Institute of Technology, Cambridge, MA 02139, USA\\
	$^3$Beijing National Laboratory for Condensed Matter Physics, Institute of Physics, Chinese Academy of Sciences, Beijing 100190, China\\
	$^4$School of Physical Sciences, University of Chinese Academy of Sciences, Beijing, 100190, China\\
	$^5$National High Magnetic Field Laboratory, MS E536, Los Alamos National Laboratory, Los Alamos, NM 87545, USA\\
	$^6$Songshan Lake Materials Laboratory, Dongguan, Guangdong, 523808, China
}

\begin{abstract}
In metals, electrons in a magnetic field undergo cyclotron motion, leading to oscillations in physical properties called quantum oscillations. This phenomenon has never been seen in a robust insulator because there are no mobile electrons. We report the first exception to this rule. We study a Mott insulator on a kagome lattice which does not order magnetically down to milli-Kelvin temperatures despite antiferromagnetic interactions. We observe a plateau at magnetization equal to
$\frac{1}{9}$ Bohr magneton per magnetic ion, accompanied by oscillations in the magnetic torque, reminiscent of quantum oscillations in metals. The temperature dependence obeys Fermi distribution. These phenomena are consistent with a quantum spin liquid state whose excitations are fermionic spinons with a Dirac-like spectrum coupled to an emergent gauge field.
\end{abstract}

\maketitle

In conventional metals, electrons form Landau Levels in a magnetic field, leading to magnetic oscillations in their physical properties. In the absence of charged Fermi surfaces, a robust insulator is NOT expected to host any quantum oscillations. Therefore, the recent observations of Landau Level quantization in narrow-gap, correlated Kondo insulators~\cite{Xiang2018, Li2020, Tan2015, Xiang2021,Senthil2019, Sodeman2018} have created a lot of excitement. These developments lead naturally to the next question: can quantum oscillations be observed in wide-gap correlated insulators, in particular, in nonmagnetic Mott insulators? In lattices with an odd number of electrons per unit cell, strong repulsion between electrons may result in a Mott insulator, where the electrons are localized on lattice sites, forming $S=\frac{1}{2}$ moments. The moments interact via antiferromagnetic (AF) interactions, and usually form an ordered AF state. In frustrated lattices, magnetic ordering may be suppressed due to frustration and quantum fluctuations, resulting in a novel state of matter called the {\it quantum spin liquid} (QSL)~\cite{Anderson1973, Anderson1987}. Some versions of the quantum spin liquid are predicted to host exotic particles such as gapless fermions which carry S=$\frac{1}{2}$  but no charge (called spinons), coupled to an emergent gauge field~\cite{Savary,Zhou}. Indeed it has been proposed that these low energy excitations can lead to quantum oscillations in certain spin liquid candidates which are charge insulators ~\cite{Motrunich}

The kagome lattice exhibits a high degree of frustration and is a strong candidate for hosting a spin liquid~\cite{Ran2007, Hermele2008,Savary, Zhou, Norman}. Experimentally,  herbertsmithite is the most famous example of a kagome Mott insulator, leading to fascinating discoveries~\cite{Han2012}. Unfortunately, while the Cu ions in the kagome layers remain pristine \cite{Smaha2020,Freedman2010}, a significant fraction of the Zn sites that lies between the kagome planes are substituted by Cu, creating impurity spins that can dominate the low temperature spectrum, thermodynamics, and magnetic properties~\cite{Vries2008, Wei2021}. A search for oscillations in this kagome Mott insulator was conducted but was unsuccessful~\cite{Asaba2014}. The recent discovery of YCu$_3$(OH)$_6$Br$_2$[Br$_{1-y}$(OH)$_y$] (YCOB), in which Zn is replaced by Y, solves the site mixing problem thanks to the very different ionic sizes of Y and Cu~\cite{JMMM, Zeng2022, Liu2022, Hong2022}. These materials do not show magnetic order down to 50 mK, but there remains disorder in the exchange constants caused by the random replacement of Br above and below the Cu hexagons by OH~\cite{Liu2022}. It should also be mentioned that the so-called perfect Y-based kagome crystal YCu$_3$(OH)$_6$Cl$_3$ with Cl instead of Br and without the OH disorder, was the first to be developed \cite{sun2016perfect} and later found to order at 15 K \cite{zorko2019coexistence}. This is not unexpected because the presence of a Dzyaloshinskii Moriya (DM) coupling is theoretically known to favor ordering \cite{Cepas2008,BernuPRB2020,Messio2010}. On the other hand, the closely related Y-kapellasite Y$_3$Cu$_9$(OH)$_{19}$Cl$_8$ has a tripled inplane unit cell and orders at a lower temperature of 2.1 K \cite{ChatterjeePRB2023}. Therefore, OH substitution and the subsequent disorder may play a role in suppression AF ordering, as discussed in a recent detailed study \cite{xu2024magnetic}. The role of disorder is a complicated question and we will defer further comments to the Discussion section below.  At this point we simply note this class of crystals exhibits a great deal of richness and complexity and we are motivated to study the magnetic behavior of YCOB under intense magnetic fields. We should also emphasize that while YCOB in zero field has been claimed to harbor Dirac spinons \cite{Zeng2022,Liu2022} and have been under intense studies \cite{xu2024magnetic, zeng2024spectral}, the current paper focuses on the behavior under strong magnetic field, and the connection of the new state of matter that we uncover with the zero field case is left for future studies.

\section*{Results}
\begin{figure*}[!htb]	
	\centering
	\includegraphics[width=\textwidth]{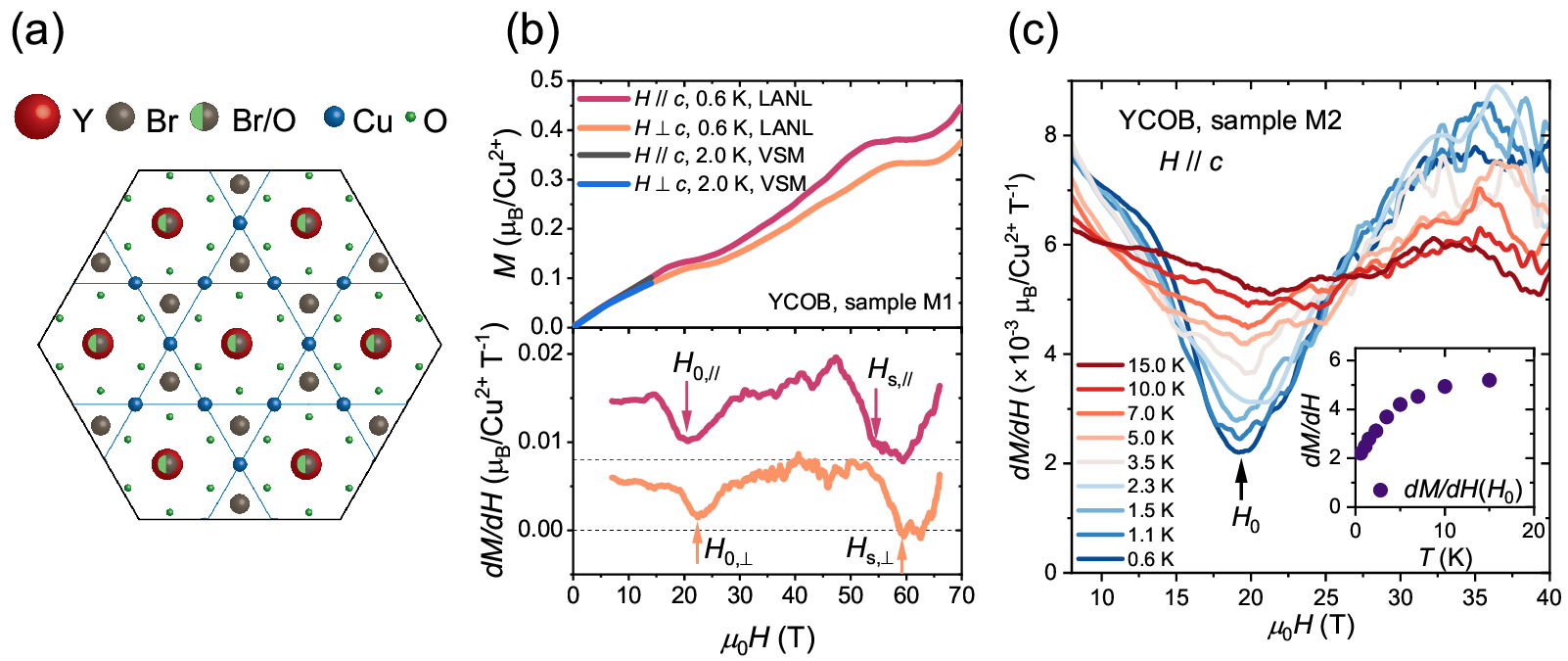}
	\caption{{\bf Magnetization plateaus.}
		(\textbf{a}) Structure of YCOB. Cu and Y are in the $z=0$ plane. The Br sites above and below Y are randomly admixed with OH. The O sites around the hexagon are alternately buckled up and down off the plane.   (\textbf{b}) Top panel: Magnetization $M$ of YCOB sample measured in pulsed magnetic fields of up to 70~T with $\textbf{H}\parallel c$ and $\textbf{H}\perp c$. The measurement setup is shown in the inset of \textcolor{blue}{SI Appendix, Fig. S2(a)}. Both $\frac{1}{9}$ and $\frac{1}{3}$ plateaus were observed in $M$. Black and blue curves show measurements in the PPMS up to 14~T in the two directions for comparison. Bottom panel: The differential magnetic susceptibility $\frac{\mathrm{d} M}{\mathrm{d} H}$ 
		measured in pulsed fields. The low-field arrow indicates the location of the Dirac point crossing $H_0$, and the high-field arrow indicates the onset of the $\frac{1}{3}$ plateau $H_{s}$. $H_0$ and $H_{s}$ along the ab-plane shift to higher fields, consistent with the known $g$-factor anisotropy.  
		For clarity, the $H\parallel c$ curve is shifted with zero marked by the dotted line. (\textbf{c}) The temperature dependence of $\frac{\mathrm{d} M}{\mathrm{d} H}$ measured on YCOB sample M2 with $H \parallel c$. Inset: the temperature dependence of $\frac{\mathrm{d} M}{\mathrm{d} H}$ at $H = H_0$, which corresponds to the minimum within the $\frac{1}{9}$ plateau region. The unit of  $\frac{\mathrm{d} M}{\mathrm{d} H}$ is $10^{-3}\mu_B/\textup{Cu}^{2+}/\textup{T}$. 
	}

	\label{Fig1}
\end{figure*}

Five YCOB single-crystal samples are used in the torque magnetometry measurements, and two batches of YCOB single crystals are measured using extraction magnetometry. 
Fig. \ref{Fig1}(a) shows the crystal structure of YCOB, with detailed information given in Ref. \cite{Zeng2022}. Samples M1 and M2 comprise several thin crystals stacked in a Vespel ampoule with their $c$-axes or $ab$-planes aligned. We use compensated-coil extraction magnetometry~\cite{Goddard2008} [see sketch in \textcolor{blue}{SI Appendix, Fig. S2(a)}] to measure the overall magnetization in pulsed magnetic fields $\mu_0H$ of up to  60~T and 73~T. The field is applied parallel and perpendicular to crystal $c$-axis in separate experiments. Sample S1 is the main single crystal measured using cantilever magnetometry in a DC magnet ($0\leq \mu_0 H \leq 41$~T). The magnetic torque of three more YCOB single crystals (Sample S2, D1, and D2) was measured in two DC magnets, and the torque of one more YCOB single crystal (Sample S5) was measured in a 60 T pulsed magnet. Details and sample growth information can be found in \textcolor{blue}{section Materials and Methods}.

The overall magnetization curves shown in the upper panel of Fig.~\ref{Fig1}(b) are derived by averaging data from multiple field pulses recorded with $T\approx 0.6$~K. A small anisotropy is observed in both $M$ and the characteristic fields; the ratio is consistent with the reported $g$-factor anisotropy~\cite{Zeng2022}. For $\mu_0 H > 55$~T, the  magnetization reaches a plateau at around $0.35 \mu_B$ per Cu atom (where $\mu_{\rm B}$ is the Bohr magneton), indicating the $\frac{1}{3}$ plateau~\cite{Schulenburg2002, Okamoto2011, Nishimoto2013, Ishikawa2015, Yoshida2022}. According to the model in Ref.~\cite{Nishimoto2013}, the $\frac{1}{3}$ plateau begins at 0.83$J$ in a kagome Heisenberg antiferromagnet, where $J$ is the nearest-neighbor AF exchange constant. We can therefore estimate that $J \approx 44.5$~K, which is comparable to the value reported in Ref.~\cite{Liu2022}. In addition to the $\frac{1}{3}$ plateau, at $\mu_0H \approx 18$~T, a feature appears in the $M(H)$ curve, with $M \approx 0.11 \mu_B$ per Cu, providing evidence for a $\frac{1}{9}$ plateau. Studies conducted in parallel to ours have also  reported this plateau \cite{jeon2024one, suetsugu2024emergent}. Note that the $\frac{1}{9}$ plateau has been anticipated by theory. Density-matrix renormalization group (DMRG) work on the Heisenberg model on the 
kagome lattice has revealed a $\frac{1}{9}$ plateau~\cite{Nishimoto2013}, considered to be an indication of exotic physics~\cite{Oshikawa97,Nishimoto2013, Yoshida2022} while tensor network methods favor a valence bond solid which triples the unit cell~\cite{Picot2016,Fang2023}. A recent variational Monte Carlo calculation based on the Gutzwiller projection of a fermionic state yields a Z3 spin liquid with energy in good agreement with the DMRG calculation \cite{He2024}.

Further information comes from the differential magnetic susceptibility $\chi \equiv \frac{\mathrm{d} M}{\mathrm{d} H}$  shown in the lower panel of Fig.~\ref{Fig1}(b);  a $V$-shape dip appears at the center of the $\frac{1}{9}$ plateau. The  dips are centered at $\mu_0H_0 \approx 20.4$~T for $H \parallel c$ and  $\mu_0 H_0 \approx 22.3$~T for $H \perp c$. These field differences are consistent with the $g$-factor anisotropy. Another batch of single crystals were prepared as Sample M2 for extraction magnetometry to get a more detailed $T$-dependence of the magnetization. The corresponding magnetic susceptibility is shown in Fig. \ref{Fig1}(c), and the raw magnetization data are given in \textcolor{blue}{SI Appendix, Fig. S2(a)}. We notice that the minimum in magnetic susceptibility around $H_0\sim 20$ T does not saturate when $T$ approaches zero, even at the lowest temperature (0.6 K$\sim 0.01 J$). To quantitatively evaluate the temperature dependence of the $\frac{1}{9}$ plateau, the value of $\frac{\mathrm{d}M}{\mathrm{d}H}$ at $H = H_0$ is plotted against $T$ in the inset of Fig.~\ref{Fig1}(c), showing a quasi-linear behavior as $T \rightarrow 0$, contrasting sharply with the fully gapped behavior observed at the $\frac{1}{3}$ plateau in other antiferromagnetic systems which can be understood in terms of magnon excitations \cite{Sheng2022, Kout2015}. The potential origin of this unconventional $\frac{1}{9}$ plateau will be explained in the Discussion section.

\begin{figure*}[!htb]
	\centering
	\includegraphics[width=\textwidth]{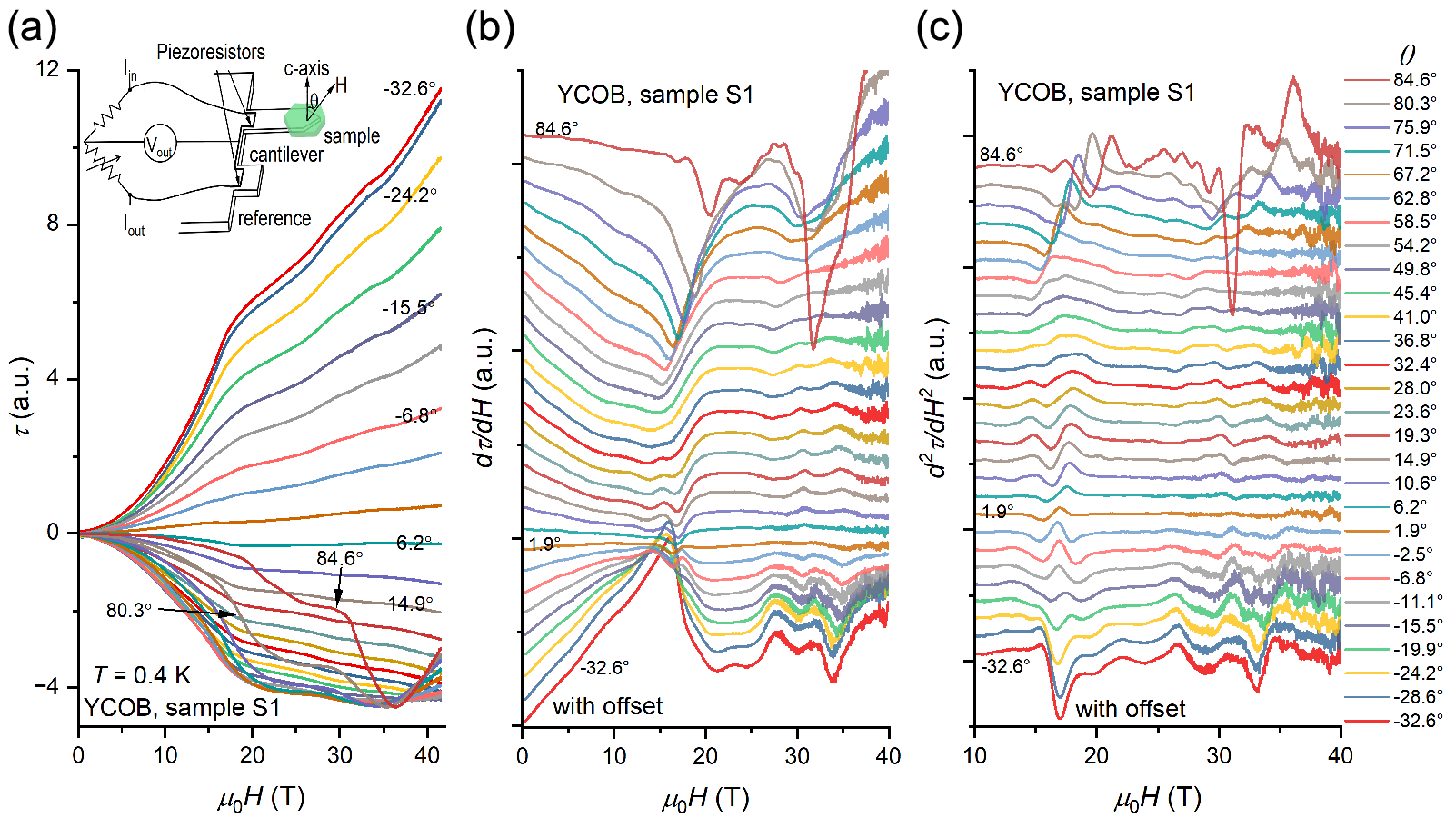}
	\caption{{\bf Oscillations in magnetic torque.}
		(\textbf{a}) Raw magnetic torque $\tau$ data, (\textbf{b}) ${\rm d}\tau/{\rm d}H$, and (\textbf{c}) ${\rm d}\tau^2/{\rm d}H^2$ versus magnetic field $H$  at $T=0.4\ $K at different angles of YCOB sample S1, measured with a piezo-resistive cantilever. $\theta$ is defined by the angle between $\vec{H}$ and $\hat{c}$. The magnetic torque was measured using a piezoresistive cantilever (setup shown in the (a) inset) in a resistive magnet up to 41~T.
	}
	
	\label{Fig_alldata}
\end{figure*}

The torque magnetometry results confirm the $\frac{1}{9}$ plateau and reveal more detail due to the finer resolution. As shown in the inset of Fig. \ref{Fig_alldata}(a), the torque magnetometry setup measures the torque due to the crystal in the direction perpendicular to the $M-H$ plane as $\tau = \mu_0 V \mathbf{M} \times \mathbf{H} = \mu_0 V M_{\rm t} H$, with $V$ the sample volume and $M_t$ the magnetization component perpendicular to $H$. In other words, instead of detecting the overall magnetization, torque magnetometry picks up the anisotropic response, either due to a higher order $H$-dependence of $M$ or off-diagonal terms of the magnetic susceptibility tensor. The angular dependence of $\tau$ measured in YCOB sample S1 is shown in Fig. \ref{Fig_alldata}(a), where $\theta$ is the angle between $\vec{H}$ and $\hat{c}$. First, the bump observed in the $M-H$ curve of Sample M1 and M2 at $\mu_0H \approx 18$~T also occurs in $\tau$.  More surprisingly, a series of dips and peaks are observed at $H> 20$~T, which are clearly seen in the derivative of torque across most angles, as shown in Fig. \ref{Fig_alldata}(b). Taking second derivatives of $\tau$ [Fig.~\ref{Fig_alldata}(c)] with respect to $H$ effectively removes any quadratic background, allowing the oscillatory pattern to be further clarified without creating new oscillations.

\begin{figure*}[!htb]
	\centering
	\includegraphics[width=\textwidth]{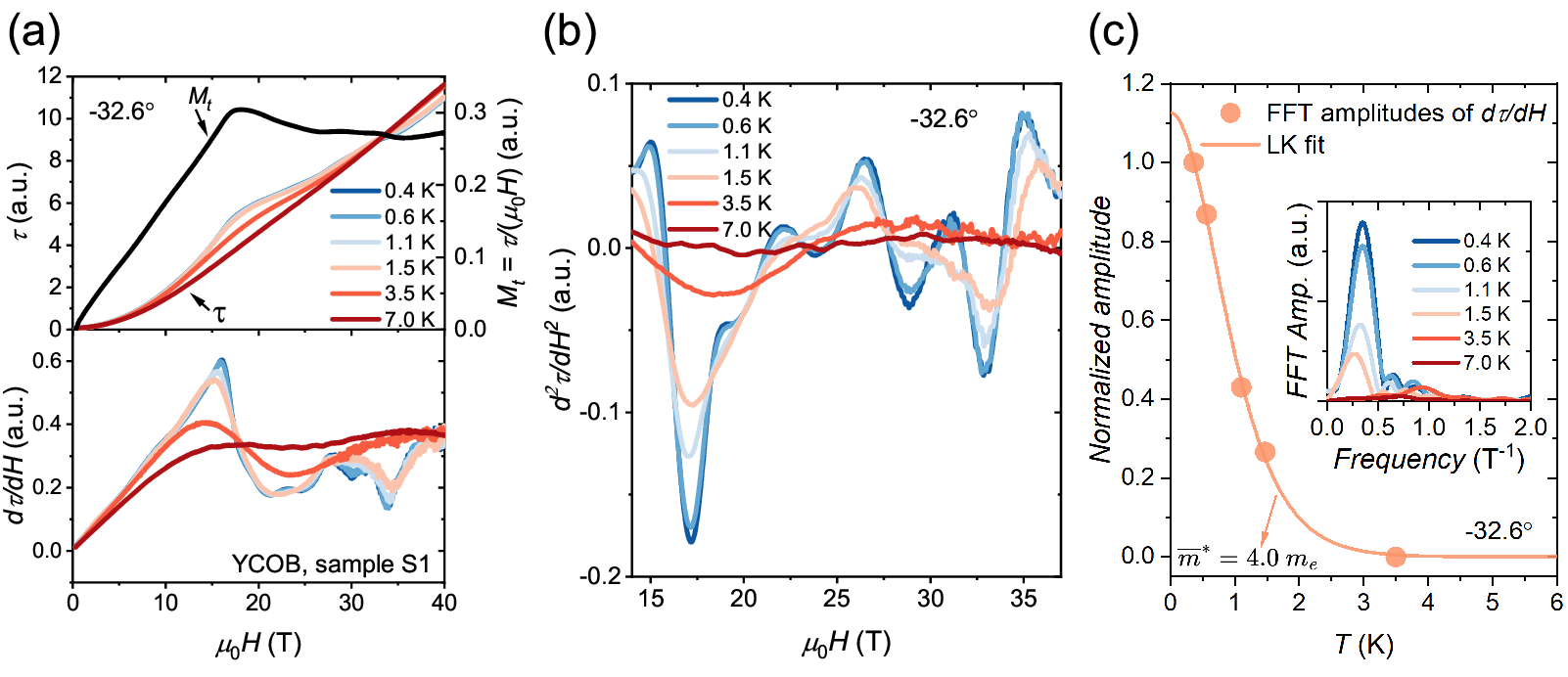}
	\caption{{\bf Temperature dependence of the magnetic oscillations.}
		(\textbf{a})  Temperature dependence of magnetic torque $\tau$ (top panel, left), ${\rm d}\tau/{\rm d}H$ (bottom panel), and (\textbf{b}) ${\rm d}^{2}\tau/{\rm d}H^{2}$ of YCOB sample S1 as a function of applied magnetic field up to 41~T at $\theta=-32.6^{\circ}$. The transverse magnetization $M_t=\tau/\mu_0H$ at $T= 0.4$~K is shown in the top panel of (a) to compare with the magnetization data. Oscillations are observed in magnetic field above the $\frac{1}{9}$ plateau at $B_{0}\approx 20$~T. The amplitude of the oscillation smears out as the temperature rises. (\textbf{c}) The normalized FFT amplitude of the ${\rm d}\tau/{\rm d}H$ oscillatory patterns after background subtraction (See \textcolor{blue}{SI Appendix, Fig. S5(b)}). The background subtraction method is shown in \textcolor{blue}{SI Appendix, Fig. S5(a)}. The solid curve shows the LK fitting, and the fitted average effective mass $\bar{m}^*$ is 4 $m_e$. Inset: The FFT analysis (See \textcolor{blue}{SI Appendix, section S5}) within FFT window [24 T, 35.7 T].
	}
	\label{Fig_Tdepend}
	
\end{figure*}

To further investigate these unusual oscillations, the $T$-dependence of $\tau$, the resulting $M_t$, the first derivate of torque $\frac{\mathrm{d}\tau}{\mathrm{d}H}$ at $\theta=-32.6^{\circ}$ are plotted in Fig.~\ref{Fig_Tdepend}(a). Similar data for two other values of $\theta$ are plotted in \textcolor{blue}{SI Appendix, Fig. S6}. The magnetic oscillations above 20 T are rapidly suppressed with increasing temperature, becoming unobservable above 3 K. In Fig.~\ref{Fig_Tdepend}(b), the second derivative of the torque is plotted to offer a clearer representation of the magnetic oscillations. The oscillations are approximately evenly spaced in $H$, in contrast to the $1/H$ periodicity of magnetic quantum oscillations in metals. As can be seen in Fig.~\ref{Fig_Tdepend}(b), as $T$ increases, the oscillation amplitude shrinks whilst the period stays the same. Figure~\ref{Fig_Tdepend}(c) shows the average amplitude of the $\frac{\mathrm{d}\tau}{\mathrm{d}H}$ oscillations, after background subtraction in the range [24 T, 35.7 T], obtained using Fast Fourier Transformation (FFT - see \textcolor{blue}{SI Appendix, section S5}), and plotted as a function of temperature $T$. The result can be fitted with the Lifshitz-Kosevich (LK) formula, which gives the $T$-dependence of amplitudes, i.e., $\delta M \propto \dfrac{aT}{\mathrm{sinh}(aT)}$, for quantum oscillations due to fermions with mass $m^*$ \cite{Shoenberg}. Using $a=\frac{2\pi^2}{e\hbar}\frac{m^*}{\mu_0 H \cos(\theta)}$, the resulting $m^*/m_e=4$ is similar to the values  $m^*/m_e=5.2$  found for another angle (\textcolor{blue}{SI Appendix, Fig. S6}). Although the data points in Fig. \ref{Fig_Tdepend}(c) can also be fitted with other forms, such as a Gaussian distribution, we opted for the LK fit because it is based on known physics and produces a mass that is consistent with that extracted from an entirely different measurement based on the Dirac spinon model discussed later. 

To confirm the observation of the oscillations, 4 additional YCOB crystals were measured in 3 different high-field magnets, with $H$ up to 41 or 45 T in DC magnetic fields and up to 60 T in a pulsed magnet. \textcolor{blue}{SI Appendix, Table S1} summarizes the limits of the samples we tested and whether the observation was confirmed. \textcolor{blue}{Section S6} and \textcolor{blue}{Fig. S10} in \textcolor{blue}{SI Appendix} present a direct comparison of the torque signals. Additionally, we conducted a control experiment by intentionally doping Cl at the Br site \cite{xu2024magnetic}. The control sample retains the same structure and the kagome lattice formed by Cu, yet shows neither the $\frac{1}{9}$ plateau nor the magnetic oscillations (see \textcolor{blue}{SI Appendix, Fig. S9}). 

\section*{Discussion}
We first summarize the experimental observations. The measurements of magnetization and magnetic torque both reveal the $\frac{1}{9}$ plateau in $H$  along both $c$-axis and the $ab$-plane. The magnetic torque measurement further reveals many oscillations above the $\frac{1}{9}$ field, whose positions are aligned consistently in the first derivative $\frac{\mathrm{d}\tau}{\mathrm{d}H}$ and the second derivative $\frac{\mathrm{d}^2\tau}{\mathrm{d}H^2}$. These oscillations are roughly evenly spaced in $H$. Furthermore, the oscillation positions shift systematically as the $H$-field is rotated from the $c$-axis to the $ab$-plane. As $T$ increases, the oscillation positions stay the same, and their amplitude shrinks. These observations call for interpretation. 

We begin with a discussion of the   1/9 plateau. The $\frac{\mathrm{d}M}{\mathrm{d}H}$ curve shown in Fig. \ref{Fig1}(c) is V-shaped and shows a linear $T$ dependence at the dip down to our lowest temperature of 0.4 K. This suggests that gapless excitations exist at the plateau.  In contrast, the 1/9 plateau reported in ref \cite{suetsugu2024emergent} is found to saturate at low temperature and has been interpreted as forming a spin gap below 1.6 K. Two possible gapped scenarios, a Z3 spin liquid state \cite{Nishimoto2013} and valence bond crystal \cite{Picot2016, Fang2023} have been proposed to explain the emergent gap. Possible explanations for these differing observations may include sample variations, as discussed in detail in Ref. \cite{xu2024magnetic}, and the temperature resolution of the magnetization measurements performed in the high field. Further experiments are needed to clarify this discrepancy.

Next, to understand the origin of magnetic oscillations, we need to rule out extrinsic effects such as metallic impurity islands inside this 3-eV gapped insulating sample. As shown in \textcolor{blue}{SI Appendix, Fig. S8} and \textcolor{blue}{Fig. S10}, five different samples grown in separate batches have been measured in two DC magnets and one pulsed magnet to confirm the reproducibility of the observations. Additionally, an impurity metallic phase would produce $1/H$ periodic magnetic oscillations, a common feature in metals, which is contrary to the quasi-$H$-periodic oscillation pattern shown in Fig. \ref{Fig_Tdepend}(b). 

Therefore, an explanation that is intrinsic to the system is needed. In the following, we review examples in the literature where structures as a function of magnetic field have been seen in magnetization data and interpreted as a series of phase transitions. We will see that these data and their interpretation are quite different from our data in their temperature and angle dependences. Next, we will discuss an interpretation based on the Dirac spinon model. We will show that a version of this model produces a simple formula (Eq. \ref{eq_dela M} and Eq. \ref{eq_DeltaB}) which accounts for the complex temperature and angle dependence with a set of internally consistent parameters.  Of course, this does not constitute proof, but at the very least, the model allows us to organize the data in a systematic way.

\subsection{Magnetic Phase Transitions}
We note that recently oscillatory behavior in the thermal conductivity has been reported in the spin liquid candidate $\alpha$-RuCl$_3$ and interpreted as quantum oscillations \cite{Czajka2021}. This interpretation has been challenged and alternative proposals involving a series of magnetic phase transitions have been made
\cite{Bruin2022,Lefranccois2023, Suetsugu2022}. We would like to point out that there is an important difference between these oscillations and our observations. In $\alpha-$RuCl$_3$ the oscillations depend on the components of the magnetic field that are in-plane, whereas our observations are sensitive to the out-of-plane field as in conventional quantum oscillations while showing additional complex variations with the tilt angle from the c-axis. Nevertheless, this debate inspired us to consider an alternative explanation based on assuming a series of magnetic transitions. Indeed, some kagome magnet systems are known to exhibit multiple plateau transitions, as observed in kagome insulators \cite{Okuma2019, Ishikawa2015} and kagome metals \cite{Zhao2020}. In these materials, the lattice might be weakly distorted by the Zeeman energy of the magnetic field, leading to different superlattice phases of spins and resulting in classical spin up-up-down (UUD) plateaus. Experimentally, these plateaus are directly revealed in the overall magnetization. As $T$ increases, these plateau transition fields usually shift, reflecting the delicate balance between the lattice distortion and thermal activation \cite{Zhao2020}. Similarly, triangular lattice quantum magnet insulators may exhibit a dominant UUD plateau, with additional states emerging in different fields \cite{Sheng2022,Hwang2012}. These plateau states are generally observed through magnetization measurements. Torque measurements on these quantum magnets \cite{Wu2022,Rao2021} resolve the transition fields marking the beginning and end of these plateau states. The angular dependence typically does not show significant shifts in the transition field due to the isotropic nature of the Zeeman energy, with any shifts arising from the weak anisotropy of the $g$-factor. On the other hand, the peaks generally shift with increasing temperature, and sometimes neighboring peaks merge \cite{Wu2022}.  

Therefore, the first possible scenario is that our observed magnetic oscillations originate from multiple plateau states, similar to those seen in other kagome magnets. In this case, one would associate each oscillation peak (or valley) with a phase transition. We would expect sharp features in the $\frac{\rm{d}M}{\rm{d}H}$ as one crosses a phase boundary, as is seen in Cs$_2$CuCl$_4$ \cite{Tokiwa2006}, or a more “rounded” transition feature in $\frac{\rm{d}M}{\rm{d}H}$, as in Cs$_2$CuBr$_4$ \cite{Fortune2009}. Our data do not show these spikes and remain sinusoidal down to the lowest temperature. Our peak positions do not show any $T$ dependence as seen in the experiments on other systems. Furthermore, since ordered phases live between the phase boundaries, one expects to observe typical phase transition signatures in the magnetization as the temperature reaches the critical temperature, with power law exponents. This is seen in Cs$_2$CuCl$_4$ \cite{Tokiwa2006}. We do not see any evidence of such phase transitions. However, it should be noted that the possibility that the sharp features are broadened by disorder in the exchange constant which is intrinsic to YCOB \cite{Liu2022} cannot be fully ruled out.  

Finally, an important shortcoming of this scenario is that it fails to explain the systematic angular dependence of the many observed oscillations. One would need to make {\it ad-hoc} assumptions about how the phase boundaries shift with the magnetic field.

\begin{figure}[!htb]	
	\centering
	\includegraphics[width=1\columnwidth]{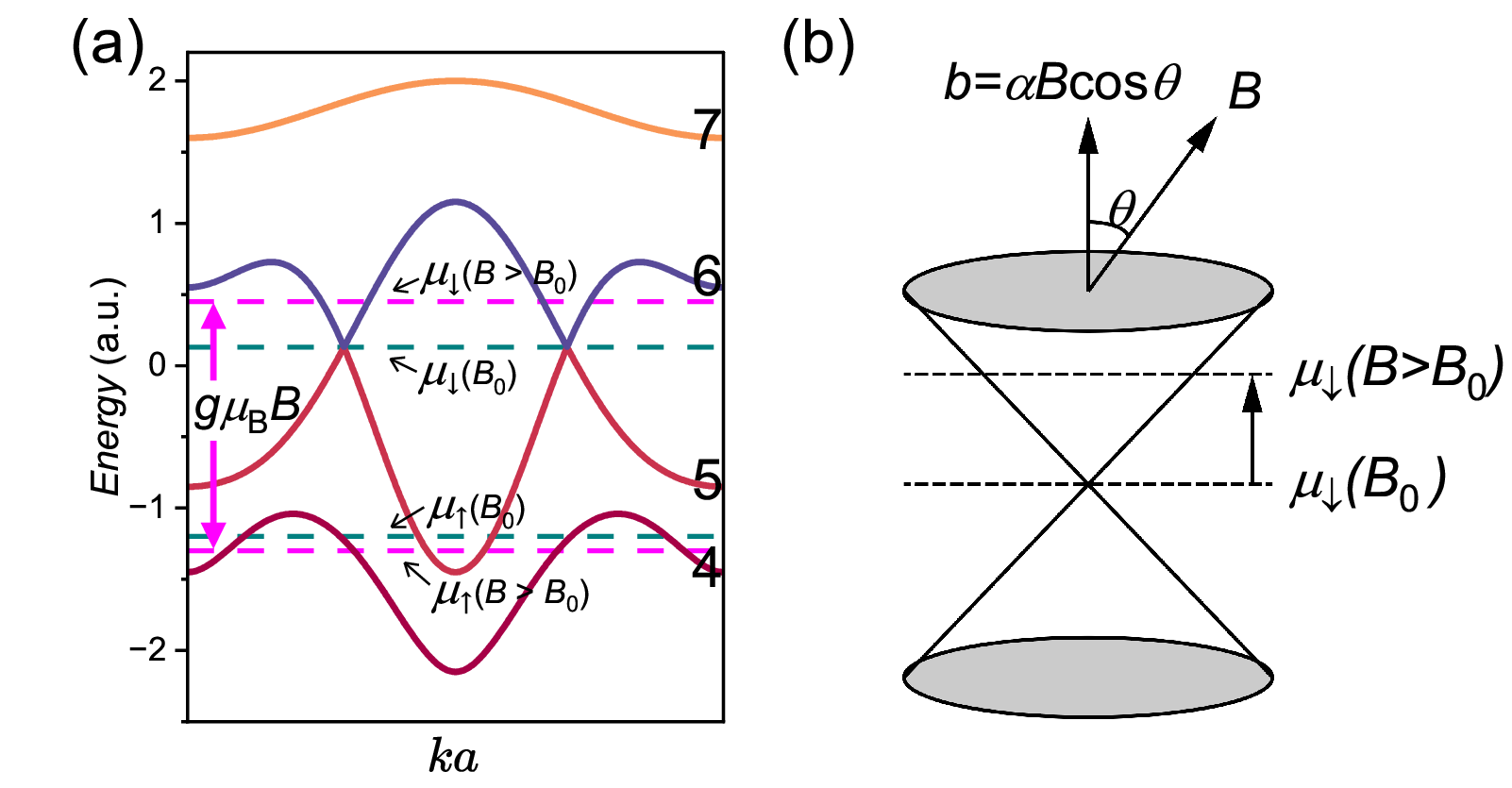}
	\caption{{\bf Dirac spinon model.}
		(\textbf{a}) A schematic band structure for bands 4-7 out of 9 bands taken along a line in k space that cuts through a Dirac node between bands 5 and 6. The chemical potentials of up and down spin near the $\frac{1}{9}$ plateau are also plotted, showing the pinning of the up spin chemical potential for $B>B_0$. (\textbf{b}) The sketch of one Dirac node around $B_0$.After applying the magnetic field, the chemical potential $\mu$ of one spin will shift upward due to the Zeeman effect. The gauge field of the spinors seen is $b=\alpha B \rm{cos}\theta$, where $\alpha$ is a coupling constant.
	} 
	
	\label{Fig4}
\end{figure}

\begin{figure*}[!htb]
	\centering
	\includegraphics[width=\textwidth]{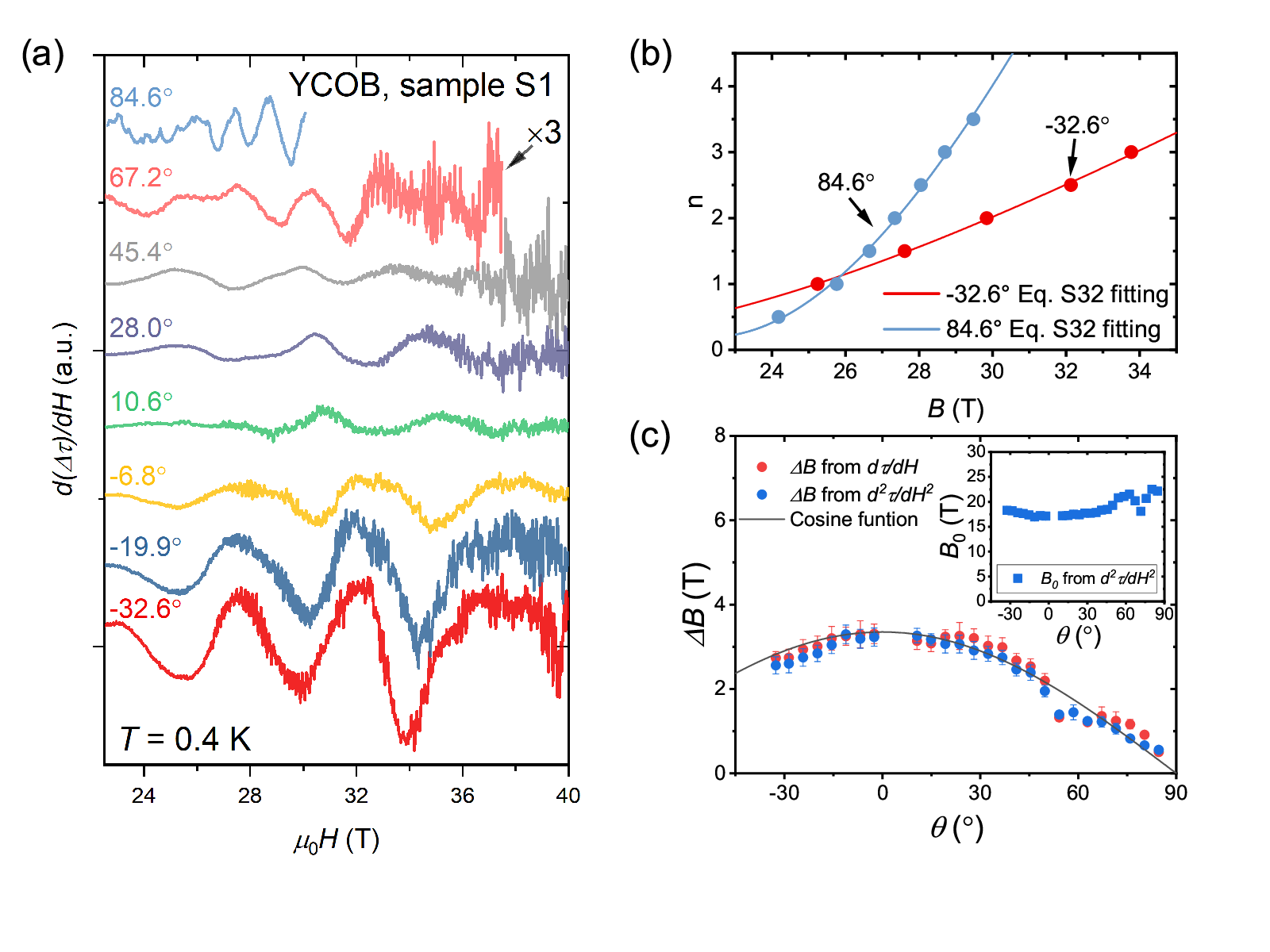}
	\caption{{\bf Angular dependence of the magnetic oscillations.}
		(\textbf{a}) The oscillations in ${\rm d}\tau/{\rm d}H$ at different angles. The oscillation component was obtained after subtraction of a Loess-smoothed background. The background subtraction method is shown in \textcolor{blue}{SI Appendix, Fig. S5(a)}. ``$\times 3$'' means the amplitudes are multiplied by 3 for clarity. 
		(\textbf{b}) Landau index plot of the oscillations at $\theta=-32.6^{\circ}$ and $84.6^{\circ}$ in (a). maximum and minimum of the oscillations are taken to be integer and half-integer values $n$ and $n+1/2$. Solid curves show the fitting results of the function $n=(B-B_{0})^{2}/(\Delta B \cdot B)-(\varphi+\pi)/2\pi$ based on \textcolor{blue}{SI Appendix, Eq. S32}.
		(\textbf{c}) The period of oscillation $\Delta B$ versus angle $\theta$. $\Delta B$ was obtained from fitting of the Landau index plots of both ${\rm d}\tau/{\rm d}H$ (red dots) and ${\rm d}^{2}\tau/{\rm d}H^{2}$ (blue dots) at different angles. The procedure is described in \textcolor{blue}{SI Appendix, section S4} and data for 12 angles are shown in \textcolor{blue}{SI Appendix, Fig. S7}. Following Eq. \ref{eq_DeltaB} a cosine fit for $\Delta B$ versus $\theta$ is shown in the black curve. Inset: The parameter $B_0$ versus angle $\theta$   from the fitting of ${\rm d}^2\tau/{\rm d}H^2$ data in \textcolor{blue}{Fig. S7}.
	}
	\label{Fig_angle}
\end{figure*}

\subsection{Dirac spinon model}
Below we describe a phenomenological fermionic model, sketched in Fig. \ref{Fig4}, which can explain the $\frac{1}{9}$ magnetization plateau, and its temperature dependence, plus features of the magnetization oscillations and their temperature dependence that we have identified so far.  

As phenomenology, we do not attempt to start with a microscopic model, but it should be mentioned that YCOB is believed to have substantial randomness in the exchange coupling which is related to the random replacement of one of the  Br sites off the kagome plane by OH \cite{Liu2022}. This replacement distorts the Cu-OH-Cu bonds, leading to an alternation of the exchange $J$ around the hexagon. This bond alternation is ordered in Y-kapellasite \cite{ChatterjeePRB2023}, leading to a tripling of the inplane unit cell and a large reduction in the ordering temperature. In YCOB  these distorted hexagons are randomly distributed and depending on their fraction  \cite{xu2024magnetic} can totally suppress AF order. It appears that this kind of correlated disorder plays an important role in countering the DM term and stabilizing a spin liquid ground state for zero or small magnetic fields. The precise magnitude and the effect of this correlated disorder is not well known. In ref. \cite{Liu2022} a variation of $J$ by as much as 70\% has beens proposed. However, this number is based on fitting a hump in the magnetic susceptibility data, and its reliability is in question because it was shown in ref \cite{xu2024magnetic} that the hump is insensitive to the fraction of hexagons and may have a different origin.  We note that neutron scattering at $B=0$ found peaks which are narrow in momentum space that disperse rapidly \cite{zeng2024spectral}, which suggests that randomness may not be playing a dominant role. Furthermore, in the presence of a large magnetic field near the 1/9 plateau, the Zeeman energy is large and the effect of disorder may be less important.  Within our picture, randomness can lead to a broadening of the fermion bands.  Given the level of uncertainty, we do not take disorder broadening into account in our phenomenological model.

We begin by addressing the $V$-shaped $\chi$ and its $T$ dependence as shown in Fig. \ref{Fig1}(c). In the case of the $\frac{1}{3}$ magnetization plateau, a $V$-shaped $\chi$ has previously been proposed to be indicative of an incompletely developed plateau \cite{Smirnov2017}, but our temperature dependence in Fig. \ref{Fig1}(c) does not support this interpretation because in that case $\chi$ typically saturates at low temperatures. This is in contrast with the approximate linear $T$ behavior at the dip minimum seen in Fig. \ref{Fig1}(c) inset. (See \textcolor{blue}{SI Appendix, Fig. S2} for a more detailed comparison.) The $V$-shaped $\chi$ and the gapless behavior motivate us to consider a fermionic model. In a model where the low energy excitations are fermions occupying energy bands, $\chi$ is proportional to the fermion density of states (DOS)  $D(E)$ at the Fermi level. Consequently, the $V$-shaped $\chi \propto |B-B_0|$ and its linear $T$ dependence shown in Fig. \ref{Fig1}(c) inset suggests that $D(E) \propto |E|$. (Since $\chi$ is small, $B\approx \mu_0H$; we define $B_0=\mu_0H_0$ and will use $B$ and $\mu_0H$ interchangeably.) This suggests the V-shape could originate from Dirac fermions. Furthermore, the temperature dependence of $\chi$ can be fitted by a Fermi-Dirac based thermal broadening model, as shown in \textcolor{blue}{SI Appendix section S1, Fig. S2(b)}, and \textcolor{blue}{Fig. S4}, which is another evidence for the fermionic picture. 

Regardless of their origin, if the excitations are fermionic, what must the band structure look like? In order to explain the $\frac{1}{9}$ plateau, there must be a gap or a Dirac node in the fermion band so that the down (up) spin bands hold 5 (4) fermions. This requires 9 bands, which can come only from a tripling of the unit cell. This is accomplished either by breaking translation symmetry, or (without doing that) by imposing $2\pi/3$ flux per unit cell. We note that the Gutzwiller projection of the latter state is precisely the Z3 spin liquid state found in Ref. \cite{He2024}. In either case, we sketch the schematic band structure of the 4th to 6th bands in Fig. \ref{Fig4}(a) (For simplicity, we assume the band dispersion is the same for up and down spin. Since the result depends only on the density of states near the Fermi levels, the result remains the same if the bands are spin-dependent.)  A $V$-shaped $\chi$ suggests a linear density of states. In order to explain the $V$-shaped $\chi$, we assume a Dirac node with velocity $v_D$ between band 5 and 6. A small gap at the Dirac node will not change our analysis. Bands 4 and 5 may be separated by a gap or may overlap. We assume the latter for reasons that will become clear later. 

At field $B=B_0$ the down spin chemical potential $\mu_\downarrow$ is at the Dirac point while the up spin chemical potential $\mu_\uparrow$ is between band 4 and 5, as shown in Fig. \ref{Fig4}(a). Upon increasing $B>B_0$, $\mu_\downarrow$ moves up while 
$\mu_\uparrow$ moves down, and their difference is given by the Zeeman splitting $g\mu_B B$. 
The differential susceptibility $\chi= \frac{\mathrm{d} M}{\mathrm{d} H}$ is given by: (see \textcolor{blue}{SI Appendix, section S1, Eq. S11})
\begin{equation}
	\chi \propto (({D(\mu_\uparrow)^{-1}+D(\mu_\downarrow)})^{-1})^{-1}
	\label{eq_diff_chi}
\end{equation}
where $D(E)$ is the density of states at energy $E$. Note that instead of adding the density of states at the up and down spin chemical potentials, we add their inverse. Hence, $ \chi$ is dominated by the \textit{smaller} of the two. To understand this rather counter-intuitive result, let us consider the situation shown in Fig. \ref{Fig4}(a) where  $\mu_\downarrow$ is near the Dirac node and has a much smaller density of states compared with $\mu_\uparrow$ which lies in the middle of bands 4 and 5. In order to fix the total fermion density, $\mu_\uparrow$ will be nearly pinned while $\mu_\downarrow$ moves up at the rate of the full Zeeman splitting  $g\mu_B B$. The differential susceptibility ${\rm d}M/{\rm d}H$ then comes mainly from the movement of $\mu_\downarrow$ and is sensitive to $D(\mu_\downarrow)$, in agreement with Eq.\ref{eq_diff_chi}. We parameterize this movement by $\mu_\downarrow= (g'/2) \mu_B$ where $g' \approx 2g$. It is satisfying that the doubling of the $g$-factor that was needed to fit the thermal smearing of the $V$-shaped $\chi$ comes out naturally, as discussed in \textcolor{blue}{SI Appendix, section S1}. If there were a gap between band 4 and 5, $\chi$ would remain zero as long as $\mu_\uparrow$ is in this gap, giving rise to a $U$-shaped $\chi$, which does not agree with the data.  

The above discussion shows that the susceptibility data alone already places considerable constraints on the nature of the fermion bands. Can the same band structure explain the quantum oscillations? Owing to the large charge gap, the fermions must be charge neutral, leading us to interpret the fermions as fractionalized spinons that are necessarily coupled to an emergent gauge field~\cite{Savary, Zhou}.  Recall that  conventional two-dimensional (2D) metals exhibit magnetization oscillation due to Landau quantization:
\begin{equation}
	M(B) \propto   
	-{\textup{sin}}(\frac{A_{FS} \phi_0}{2 \pi B \cos(\theta)} -\varphi).
	\label{eq_metal M}
\end{equation}
where $A_{FS}$ is the Fermi surface area and $\phi_0=h/e$.

The application of this equation to our problem requires several modifications. First, for $B>B_0$ a Fermi surface is formed in the down spin band with an area $A_{FS}=\pi k_F ^2$ where $k_F=\frac{g^{\prime}\mu_B}{2 \hbar v_D} (B-B_0) $. Second, spinons with a Fermi surface are coupled to a U(1) gauge field, which is the 2D analog of the electromagnetic field in our world~\cite{Savary, Zhou, Motrunich}. Because it is 2D, the gauge magnetic field $b$ is always perpendicular to the plane and produces Landau levels in the spinon spectrum. It is useful to introduce the parameter $\alpha=b/(B \cos(\theta))$ to characterize the relative strength of the gauge field and  $B$. Then we can simply replace $B$ in the denominator of Eq.\ref{eq_metal M} by $|\alpha| B$. The origin of  $b$ and its relation to the external field $B$ will be discussed later.

Putting everything together, for $kT<< \mu_\downarrow$ the oscillatory part of the torque  per area $L^2$  is given by (See \textcolor{blue}{SI Appendix, section S3})
\begin{equation}
	\frac{\tau}{BL^2} = -\textup{sin}\theta C_0 (B-B_0) \frac{aT}{\sinh{aT}} {\textup{sin}}(2\pi \frac{(B-B_0)^2}{B \cdot \Delta B} -\varphi).
	\label{eq_dela M}
\end{equation}
where $C_0=|\alpha|\frac{e}{\pi^2 \hbar} g'\mu_B$  
and  
\begin{equation}
	\Delta B = 2 e\hbar \frac{|\alpha| v_D^2}{((g'/2)\mu_B)^2} \rm{cos}\theta.
	\label{eq_DeltaB}
\end{equation}
It is important to note that $B$ plays a dual role. Due to the Zeeman effect, it gives rise to a Fermi surface for the down spin band, leading to
the $(B-B_0)^2$ factor inside the sine in Eq. \ref{eq_dela M}. This $B$ dependence is isotropic apart from the small $g$ factor anisotropy.
On the other hand, $B$ also produces a finite gauge field $b$, which depends on the component of the $B$-field perpendicular to the plane. This gives rise to quite complex behavior as a function of $B$-field and its angle $\theta$. As we will see, a single equation (\ref{eq_dela M}) captures the complexity of the data as well as their temperature dependence.  

Next, we compare the experimental results with predictions from the Dirac spinon model. Figure~\ref{Fig_angle}(a) presents the $\theta-$dependence of the magnetic oscillations, and the background subtraction method is given in \textcolor{blue}{SI Appendix, Fig. S5(a)}. The roughly even spacing of the magnetic oscillations persists over a wide range of $\theta$. Moreover,  as $\theta$ changes, the oscillation peaks and dips gradually shift, and the intervals between them progressively decrease. This suggests an orbital origin of the oscillations. Together with the Lifshitz-Kosevich temperature dependence, we are led to attempt a Landau Level indexing as a function of $B$ based on Eq. \ref{eq_dela M} (See \textcolor{blue}{SI Appendix, Eq. S32} for the Landau Level indexing). The result is shown in Figure \ref{Fig_angle}(b) for two angles, using the data from Fig. \ref{Fig_angle}(a) (more plots for 12 angles are shown in \textcolor{blue}{SI Appendix, Fig. S7}). At $B$ much larger than $B_0 \approx 20$~T, the indexing plot is quite linear, in agreement with the above-mentioned rough $B$-periodic pattern. However, as $B$ gets closer to $B_0$, the curves become nonlinear. The nonlinear dependence of the Landau-Level index as a function of $1/B$ is also displayed in \textcolor{blue}{SI Appendix, Fig. S5(d)}; such behavior contrasts sharply with that of conventional quantum oscillations in metals.

Fig. \ref{Fig_angle}(c) displays results from conducting Landau index fitting for all the measured tilt angles using \textcolor{blue}{Eq. S32} and \textcolor{blue}{Eq. S34} in \textcolor{blue}{SI Appendix}. The procedure to obtain $\Delta B$ and the corresponding uncertainty are discussed in \textcolor{blue}{SI Appendix, section S4}. The obtainted $\Delta B$ values from $\frac{\mathrm{d}\tau}{\mathrm{d}H}$ (red dots) and $\frac{\mathrm{d}^2\tau}{\mathrm{d}H}^2$ (lue dots) are consistent with each other.
Notably, $\Delta B$ shows a large change which follows closely the  $\cos(\theta)$ dependence predicted by Eq. \ref{eq_DeltaB}. This is a strong indication that the oscillations depend on the $c$-component of the magnetic field and that they have an orbital origin. 

We have fitted the temperature-dependent data using the standard 2D LK form for a parabolic band with mass $m^*$.  The Dirac spectrum requires a different expression for $a$ ~\cite{Igor2011, Thesis}. However, it can be cast in terms of $m^*/m_e$
with the following relation:
\begin{equation}
	\frac{m^*}{m_e}=\frac{(g'/2)\mu_B(B-B_0)}{|\alpha| v_D^2 m_e}.
	\label{eq_mstar}
\end{equation}
Note that $\frac{m^*}{m_e}$ is independent of the angle $\theta$ and proportional to $B-B_0$. In Fig. \ref{Fig_Tdepend}(c) we find an average  value of $\frac{m^*}{m_e}\approx$~4. The prediction that $m^*$ increases with increasing $B$ as  $B-B_0 $ is also in rough agreement with the LK fit of the amplitude at fixed $B$ shown in \textcolor{blue}{SI Appendix, Fig. S5(c)}. We note that Eq. \ref{eq_DeltaB} and \ref{eq_mstar}, depend on a single unknown parameter $|\alpha| v_D^2$. Using the measured value of $\Delta B$, We can use Eq. \ref{eq_DeltaB} to extract the product $\sqrt{|\alpha|}v_{\rm D}$. Taking $g'=2g\approx 4$ we find $\sqrt{|\alpha|} v_{\rm D} \approx 39~$meV$\cdot$\AA$\approx 5.89\times 10^3$~m/sec. Plugging this into Eq. \ref{eq_mstar} and using $B-B_0=10$ T, we find $m^*/m=5.78$,  in reasonable agreement with the value 4 from the LK fit.  

Since $m^*$  is extracted from the temperature dependence and $\Delta B$ from the period, the agreement of the value of $m^*$ extracted from two independently measured quantities serves as a stringent test of the model of fermions with Landau level quantization; its validation provides key support for this picture. 

\section*{Origin of the gauge magnetic field.}
Now, we address a key question: In the spinon model, what is the origin of the gauge magnetic field $b$, and how is it related to the applied field? The literature has discussed two possibilities. This discussion is quite general and aplies equally well to spinon with any dispersion, whether it is Dirac or not.

\noindent
1. The spinon-chargon mechanism.  Instead of a spin Hamiltonian, we go back to the Hubbard model with finite onsite repulsion $U$ and hopping $t$, which includes charge fluctuations described by a gapped boson called chargon~\cite{LeeSS}. Even though the spinon is charge neutral, its current is accompanied by a back-flow of orbital chargon current, which couples to the external $B$ field. In this way, $b$ is proportional to the $c$-component of the magnetic field $B\cos(\theta)$. (See \textcolor{blue}{SI Appendix, section S9}.)  A physical way to view this effect in spin language was given by Motrunich for a triangular lattice~\cite{Motrunich}. He showed that there is a linear coupling between the physical magnetic flux through a triangle on-site $i,j,k$ and the scalar chirality $C$ = $\langle\vec{S}_i \cdot (\vec{S}_j\times \vec{S}_k)\rangle$. Wen, Wilczek, and Zee~\cite{Wen} showed that a spinon hopping around the triangle picks up a Berry phase given by $C$, which is interpreted as the gauge flux through the triangle. Hence $b \propto C\propto B \cos(\theta)$.  Near the Mott transition, the ratio $\alpha=\frac{b}{B\cos(\theta)}$ is of order unity ~\cite{Motrunich} but is expected to diminish as the Mott band gap increases. 

\noindent
2. The DM mechanism. In a recent paper \cite{Kang}, it is pointed out the Dzyaloshinskii-Moriya (DM) interaction,  which is known to exist in these kagome compounds due to the non-collinear Cu-O-Cu bond (see Fig. \ref{Fig1}(a)), can give rise to an effective gauge field. The idea is that even in the absence of spin ordering, the DM term causes a canting of the nearest neighbor spins on sites $i,j$, so that $\langle (\vec{S}_j\times \vec{S}_k) \rangle$ has an average $z$ component.
In the presence of spin polarization $\langle\vec{S}_z\rangle$ is also nonzero. As a result, the scalar chirality on three sites that form a triangle $C=\langle\vec{S}_i \cdot (\vec{S}_j\times \vec{S}_k)\rangle$ is finite and $\propto \langle\vec{S}_z\rangle$ after a mean-field factorization. Since $b$ is proportional to the chirality $C$~\cite{Wen}, we write  $b=\gamma \langle\vec{S}_z\rangle $ where $\gamma$ is a constant. Note that $b$ is again proportional to $\cos(\theta)$. In Ref.~\cite{Kang}, it was shown that this mechanism generates a gauge magnetic field which is equivalent to 120~T in a direction opposed to the external $B$, hence $\alpha \approx -4$. Note that
this description deals only with spins and remains valid in a large band gap insulator. Its origin is spin-orbit coupling, which makes the orbital nature of the effect physically more transparent. 

In the spinon-chargon mechanism $\alpha$ is a constant but in the DM mechanism, $|\alpha|=\gamma \langle|\vec{S}|\rangle /B$ and can have some $B$ dependence if $|\langle\vec{S}\rangle|$ is not linear in B, as is the case near the $\frac{1}{9}$ plateau. However, this dependence is less than 12\% in the relevant $B$ range between 20~T and 40~T, as shown in the inset of \textcolor{blue}{SI Appendix, Fig. S11(a)}. \textcolor{blue}{Fig. S11(b)} shows the data fitted using the DM mechanism; the fit quality and the resulting parameters are similar to those from the spinon-chargon mechanism. Below, we will continue our discussion assuming a constant $\alpha$.
Recall that we use Eq. \ref{eq_DeltaB} to extract  $\sqrt{|\alpha|} v_{\rm D} \approx 39~$meV$\cdot$\AA. 
The velocity itself can be extracted from the slope of $\chi$ vs $B$, yielding the average of $v_{\rm D} \approx 29.3 \sqrt{M_{\downarrow}/9}~$meV$\cdot$\AA, where $M_{\downarrow}$ is the number of Dirac nodes in the original Brillouin zone which is multiple of 3 due to unit cell tripling (\textcolor{blue}{SI Appendix, section S8}). In SI section S1 we use the linear $T$ term in Fig. 1c inset to extract a very similar value of $v_{\rm D} \approx$ 34.1$\sqrt{M_{\downarrow}/9}~$meV$\cdot$\AA. We  find that $|\alpha|$ is $\approx$ 1.33 and 2.3  if $M_{\downarrow}$ is 9 or 3 respectively.  In the DM mechanism, it is estimated that $|\alpha| \approx 4$ ~\cite{Kang} while in the spinon-chargon mechanism,  $\alpha \approx t/U$,  which would be somewhat less than unity (\textcolor{blue}{SI Appendix, section S9}). Thus our analysis tends to favor the DM mechanism. 

\section*{Summary and Conclusions}

In summary, we report the discovery of a $\frac{1}{9}$ magnetization plateau in the Mott insulator YCOB under an intense magnetic field. Right above this field, magnetic oscillations appear in the magnetic torque of this robust insulator with a charge gap of 3 eV. The data are well reproduced in several samples (\textcolor{blue}{SI Appendix, Fig. S10}, detailed discussion in \textcolor{blue}{section S6}). These oscillations are characterized by the following key features. (1)~The oscillations are roughly periodic in large applied fields $B$. (2)~Their period, $\Delta B$, follows a $\cos \theta$ dependence, suggesting that they are caused by an orbital effect. (3)~The $T$ dependence of the oscillation amplitude is consistent with the LK formula. Point (1) and the fact that the oscillations are tied to the  $\frac{1}{9}$ plateau rule out trivial explanations such as small metallic inclusions. Our data do not resemble structures in the magnetization curves that have been reported up to now, which have been interpreted based on conventional pictures of multiple magnetic phase transitions. In contrast, a phenomenological Dirac spinon model based on the spin liquid picture was successful in organizing the complex evolution of the oscillations as a function of field, angle, and temperature.

\section*{Materials and Methods}

Single crystals of $\text{YCu}_3(\text{OH})_6\text{Br}_2[\text{Br}_{1-y}(\text{OH})_y]$ (YCOB) were grown using the hydrothermal method and ultrasonically cleaned in water before measurements to remove the possible impurities as reported previously \cite{Schulenburg2002, Zeng2022}. The deuterated single crystals (YCOB-D, sample D1 and D2) were synthesized using the same method with the corresponding deuterated starting materials and heavy water. The OH content of these samples (S1, S2, S5, M1, and M2) is given by $y=0.67$. The Cl-doped single crystals (YCOB-58\%Cl)) were synthesized using the same method as YCOB with 58$\%$ of the Br atoms replaced by Cl atoms, and $y$ is estimated to be 0.03. These values have been determined by single crystal X-ray diffraction. YCOB samples S1, S2, S5, and M1 come from the same growth batch, while sample M2 comes from another batch. The measurement conditions and observations of all samples are listed in \textcolor{blue}{SI Appendix, Table S1}. 

Magnetization measurements at low field ($<$ 14 T) were carried out in a Quantum Design physical property measurement system (PPMS Dynacool-14T) using the Vibrating Sample Magnetometer (VSM) option.

Magnetization measurements on YCOB M1 and M2 at high field were using a compensated-coil spectrometer \cite{Goddard2007, Goddard2008} as drawn in the inset of \textcolor{blue}{SI Appendix, Fig. S2(a)} which were performed at 65 T and 73 T pulsed field magnet at the National High Magnetic Field Laboratory (NHMFL), Los Alamos. YCOB M1 was stacked in a Vespel ampoule with $c$-axis aligned first and then restacked with $ab$-plane aligned to apply the magnetic field in the $ab$-plane.

Magnetic torque measurements on YCOB S1, S2, and YCOB-D D1 were using a piezo-resistive cantilever as drawn in Fig. \ref{Fig_alldata}(a) performed in 41 T Cell 6 (YCOB S1 and YCOB-D D1) and 45 T Hybrid (YCOB S2) DC field magnets in NHMFL, Tallahassee. The unloaded cantilever setup signals are measured in PPMS as shown in \textcolor{blue}{SI Appendix, Fig. S1(a)}. The angles of YCOB S1 measured in Cell 6 are adjusted by comparing the signal measured in PPMS under the same condition as shown in \textcolor{blue}{SI Appendix, Fig. S1(b)}. Note that the torque signals have larger positive peaks, which is caused by the asymmetric response of the piezo-cantilever to tension and compression. This has an effect on the quantum oscillation amplitude as a function of angle, but will not affect the analysis of the period or the temperature dependence at fixed angle.

Magnetic torque measurements on YCOB S5 were using a piezo-resistive cantilever performed in the 55 T Mid-Pulse magnet in NHMFL, Los Alamos.    

Magnetic torque measurements on YCOB-D D2 were using a capacitive cantilever performed in 41 T Cell 6  DC field magnets in NHMFL, Tallahassee.

The expression of Landau index fitting in Fig. \ref{Fig_angle}(b) is \textcolor{blue}{SI Appendix, Eq. S32}: $n=(B-B_0)^2/(\Delta B \cdot B)-(\varphi+\pi)/2\pi$, where $B_0=17.78$ T, $\Delta B=2.74$ T, and $\varphi=-1.4\pi$ for -32.6$^{\circ}$; while $B_0=22.40$ T, $\Delta B=0.50$ T, and $\varphi=-1.4\pi$ for 84.6$^{\circ}$.

The expression of cosine fitting in Fig. \ref{Fig_angle}(c) is $\Delta B(\theta)=\Delta B(0)\rm{cos}\theta$, and the fitted result is $\Delta B(0)=3.35$~T.

\subsection*{Data Availability}
All study data are included in the article and/or \textcolor{blue}{SI Appendix}.

{\bf Acknowledgement}
The work at the University of Michigan is supported by the Department of Energy under Award No. DE-SC0020184 (magnetization measurements) to Guoxin Zheng, Yuan Zhu, Kuan-Wen Chen, Kaila Jenkins, Aaron Chan, and Lu Li. A portion of this work was performed at the National High Magnetic Field Laboratory (NHMFL), which is supported by National Science Foundation Cooperative Agreement Nos. DMR-1644779 and DMR-2128556 and the Department of Energy (DOE).  J.S. acknowledges support from the DOE BES program “Science at 100 T,” which permitted the design and construction of much of the specialized equipment used in the high-field studies. The work at IOP China is supported for the crystal growth, by the National Key Research and Development Program of China (Grants 2022YFA1403400, No. 2021YFA1400401), the K. C. Wong Education Foundation (Grants No. GJTD-2020-01), the Strategic Priority Research Program (B) of the Chinese Academy of Sciences (Grants No. XDB33000000). The experiment in NHMFL is funded in part by a QuantEmX grant from ICAM and the Gordon and Betty Moore Foundation through Grant No. GBMF5305 to Kuan-Wen Chen, Dechen Zhang, Guoxin Zheng, Aaron Chan, Yuan Zhu, and Kaila Jenkins. P.L. acknowledges the support by DOE office of Basic Sciences Grant No. DE-FG02-03ER46076 (theory)

\vfill
\thefootnote{$\dagger$}{These authors contributed equally and shared the first authorship.}

\bibliographystyle{unsrt}
\bibliography{YCOB_Osc_v3}{}

\newpage
\section*{Supporting Information}

\renewcommand{\theequation}{S\arabic{equation}}
\setcounter{equation}{0}
\renewcommand{\thetable}{S\arabic{table}}
\setcounter{table}{0}
\renewcommand{\thepage}{S\arabic{page}}
\setcounter{page}{1}
\setcounter{subsection}{0}
\setcounter{section}{0}

\renewcommand{\thesection}{S\arabic{section}}

\let\oldthefigure\thefigure
\renewcommand{\thefigure}{S\oldthefigure}
\setcounter{figure}{0}

\section{Temperature evolution in Dirac spinon model}\label{sec_2gsmear}	

Here, we start from spin-1/2 Hamiltonian induced by an applied magnetic field:
\begin{align}
H=\sum_{<i,j>}J\vec{S_i}\cdot \vec{S_j}-g\mu_B \vec{B}\cdot \sum_{i}\vec{S_i}
\end{align}
Here we choose a uniform $J$ to describe the nearest neighbor exchange energy for simplicity. The bond randomness has been discussed in Ref. \cite{Liu2022}. For one spin, the Zeeman energy is
\begin{align}
	H_{Bi} & =-g\mu_B\vec{S_i} \cdot \vec{B} \nonumber \\ 
	& = \begin{cases} +\frac{1}{2}g\mu_B B, \ \rm{for\  up}(\uparrow) \ \rm{spin} \\-\frac{1}{2}g\mu_B B,\  \rm{for\  down}(\downarrow) \ \rm{spin} \end{cases}   
	\label{eq_muB_1}
\end{align}
The energy for up spin is $\epsilon_{\uparrow}-\mu = \epsilon+\frac{1}{2}g\mu_B B-\mu =\epsilon -\mu_{\uparrow}$. Here $\epsilon$ is the energy without magnetic field, and $\mu$ is the chemical potential, and we define the chemical potential for up spin as $\mu_{\uparrow}=\mu-\frac{1}{2}g\mu_B B$. Similarly, the energy for down spin is $\epsilon_{\downarrow}-\mu=\epsilon-\mu_{\downarrow}$, where $\mu_{\downarrow}=\mu+\frac{1}{2}g\mu_B B$. Importantly, we need to determine how $\mu$  depends on field B by enforcing the constraint $n_{\uparrow}+n_{\downarrow}$=constant, where $n_{\uparrow}$ ($n_{\downarrow}$) is the number of up (down) spins.
Start from the definition of $n_{\downarrow}$,
\begin{align}
	n_{\downarrow} = \int f_{\rm{F}-\rm{D}}(\epsilon-\mu_{\downarrow})D(\epsilon) d\epsilon
\end{align}
Here $f_{\rm{F}-\rm{D}}$ is the Fermi-Dirac distribution function. Then,
\begin{align}
	\frac{dn_{\downarrow}}{dB} = - \int \frac{\partial f_{\rm{F}-\rm{D}}(\epsilon-\mu_{\downarrow})}{\partial \epsilon} (\frac{d \mu}{d B}+\frac{1}{2}g\mu_B) D(\epsilon) d\epsilon
	\label{eq_muB_dndBint}
\end{align}
When $k_B T\ll\mu$, 
\begin{align}
	\frac{dn_{\downarrow}}{dB}=(\frac{d\mu}{dB}+\frac{1}{2}g\mu_B)D(\mu_{\downarrow})
	\label{eq_muB_dndBdown}
\end{align}
Similarly, we find
\begin{align}
	\frac{dn_{\uparrow}}{dB}=(\frac{d\mu}{dB}-\frac{1}{2}g\mu_B)D(\mu_{\uparrow})
	\label{eq_muB_dndBup}
\end{align}
Adding these equations, we use the constraint to show that the left-hand side is zero to obtain
\begin{align}
	\frac{d\mu}{dB}=\frac{1}{2}g\mu_B \frac{D(\mu_{\uparrow})-D(\mu_{\downarrow})}{D(\mu_{\uparrow})+D(\mu_{\downarrow})}
	\label{eq_muB_dudB}
\end{align}
Therefore, we have
\begin{align}
	\frac{d\mu_{\downarrow}}{dB}=\frac{1}{2}g\mu_B\frac{2D(\mu_{\uparrow})}{D(\mu_{\uparrow})+D(\mu_{\downarrow})}
	\label{eq_muB_dmudBdown} 
\end{align}
A similar equation is derived for $\mu_\uparrow$.
\begin{align}
	\frac{d\mu_{\uparrow}}{dB}=-\frac{1}{2}g\mu_B\frac{2D(\mu_{\downarrow})}{D(\mu_{\uparrow})+D(\mu_{\downarrow})}
	\label{eq_muB_dmudBup}
\end{align}
The  magnetic susceptibility is given by the imbalance between up and down spins:
\begin{align}
	\chi &=\frac{1}{2} g\mu_B (\frac{dn_{\downarrow}}{dB}-\frac{dn_{\uparrow}}{dB}) \\
	&=(\frac{1}{2}g)^2 \mu_B^2\frac{4D(\mu_{\downarrow})D(\mu_{\uparrow})}{D(\mu_{\uparrow})+D(\mu_{\downarrow})}
	\label{eq_muB_chi}
\end{align}
This gives Eq. 1 in the main text. We assume that band 5 and 6 overlap so that $\mu_\uparrow$ lies in the band and  $D(\mu_{\uparrow})$ has a finite value. On the other hand, when $B $ is near $B_0$, $\mu_{\downarrow}$ is near the Dirac point,  $D(\mu_{\downarrow})$ is much smaller than $D(\mu_{\uparrow})$. 
The RHS of Eq. \ref{eq_muB_dmudBdown} is approximately constant and we integrate Eq. \ref{eq_muB_dmudBdown} obtain
\begin{align}
	\mu_{\downarrow}  = \frac{1}{2}g^{\prime}\mu_B ( B-B_0).
	\label{eq_muB_mudown}
\end{align}
where we define the effective $g$-factor as
\begin{align}
	g^{\prime}=( 1+\frac{D(\mu_{\uparrow})}{D(\mu_{\uparrow})+D(\mu_{\downarrow})})g
\end{align}
Near the Dirac point, $D(\mu_{\downarrow})<<D(\mu_{\uparrow})$,  $g^{\prime} \approx 2g$ and $\mu_{\downarrow}$ moves twice as fast as $\frac{1}{2}g\mu_B B$.

As before, we assume $\mu_\uparrow$ lies between bands 4 and 5, so $D(\mu_\uparrow)$ is larger than $D(\mu_\downarrow)$, which lies near the Dirac node. It is given by $D(E)=\nu |E|$ where $\nu=M_\downarrow/2\pi v_D^2$ and $M_\downarrow$ is the number of Dirac nodes. Setting $D(\mu_\downarrow)=D(E=(g'/2)\mu_B(B-B_0))$ and integrating 
Eq. \ref{eq_muB_dmudBup}, we find 
\begin{equation}
	\mu_\uparrow=(g'/2)(g/2) \mu_B^24\nu(B-B_0)^2/D(\mu_\uparrow)
	\label{eq_mu_uparrow2}
\end{equation}
The important point is that $\mu_\uparrow$ moves slowly as $(B-B_0)^2$. We will make use of this in a later section on the quantum oscillation from the up-spin Fermi surface.

Next we derive the temperature smearing effect in magnetic properties. Starting from Eq. \ref{eq_muB_dndBint}, we can define
\begin{align}
	\tilde{D}(\mu_{\downarrow(\uparrow)})= -\int \frac{\partial f_{\rm{F}-\rm{D}}(\epsilon-\mu_{\downarrow(\uparrow)})}{\partial \epsilon}D(\epsilon)d\epsilon
\end{align}
At zero temperature $\tilde{D}(\mu_{\downarrow(\uparrow)})$ goes to ${D}(\mu_{\downarrow(\uparrow)})$.
The definition of magnetic susceptibility is related to the imbalanced numbers between up and down spins:
\begin{align}
	\chi &=\frac{1}{2} g\mu_B (\frac{dn_{\downarrow}}{dB}-\frac{dn_{\uparrow}}{dB}) \\
	&=(\frac{1}{2}g)^2\mu_B^2\frac{4\tilde{D}(\mu_{\downarrow})\tilde{D}(\mu_{\uparrow})}{\tilde{D}(\mu_{\downarrow})+\tilde{D}(\mu{\uparrow})}
	\label{eq_muB_chi_zero}
\end{align}
When $\mu_{\downarrow}$ is close to the Dirac point, $\tilde{D}(\mu_{\downarrow}) <<\tilde{D}(\mu_{\uparrow})$ , so
\begin{align}
	\chi \approx 4(\frac{1}{2} g\mu_B)^2\tilde{D}(\mu_{\downarrow})
\end{align}
In the $T \to 0$ limit, $\chi_0 \equiv \chi(T=0) \approx 4(g/2)^2\mu_B^2 D(\mu_{\downarrow})$ and show the V shape dip. Note this expression for $\chi$ is a factor of two larger than what one may naively expect based on the standard expression for Pauli susceptibility. This is because the spin down  chemical potential is moving twice as fast with $B$, since $g'=2g$.  At finite $T$:
\begin{align}
	\chi(T)=-\int \frac{\partial f_{\rm{F}-\rm{D}}(\epsilon-\mu_{\downarrow})}{\partial \epsilon}\chi_0(\epsilon) d\epsilon
	\label{eq_muB_chiT}
\end{align}
After plugging $\epsilon=\frac{1}{2}g^{\prime}\mu_B(B^{\prime}-B_c)$ and $\mu_{\downarrow}=\frac{1}{2}g^{\prime}\mu_B(B-B_0)$ into Eq. \ref{eq_muB_chiT}, we obtain
\begin{align}
	\chi(T,B) &=-\int \frac{\partial f_{\rm{F}-\rm{D}}[\frac{1}{2}g^{\prime}\mu_B(B^{\prime}-B)]}{\partial B^{\prime}} \chi_0(B^{\prime}) dB^{\prime}
	\label{eq_muB_smear}
\end{align}	
Eq. \ref{eq_muB_smear} is the final equation we derive to study the temperature smearing effect. In our case, around Dirac point $B=B_0$, we can take $g^{\prime}=2g$, and take the lowest temperature $\chi(T,B)$ data as $\chi_0(B)$. Fig. \ref{FigS_tor_smear} and Fig. \ref{FigS_toranalysis}(b) are the smearing results we got by using Eq. \ref{eq_muB_smear}.

Fig.~\ref{FigS_tor_smear} shows the nominal differential susceptibility of the transverse magnetization $M_t \approx \frac{\tau}{\mu_0 H}$ at $-32.6^\circ$ in YCOB sample S1. As $T$ increases to 3.5~K and 7~K, the sharp dip is smeared by thermal broadening. This behavior is quantitatively reproduced by applying Eq. \ref{eq_muB_smear}, which takes the lowest temperature data as $\chi_0(B)$. This demonstrates the fermionic origin of the magnetic susceptibility near the $\frac{1}{9}$ plateau. The fitting requires a Zeeman energy shift given by $E = (g'/2)\mu_B (B-B_0)$, which yields  $g^\prime=2g\approx 4$. 

We can consider the case $B=B_0$ at low temperature. Using the expression for $\chi_0$ and $D(\epsilon)=(M_\downarrow/2\pi \hbar^2v_D^2)|\epsilon|$ where $M_\downarrow $ is the number of Dirac nodes with velocity $v_D$, Eq. \ref{eq_muB_chiT} can be evaluated to obtain

\begin{align}
	\chi(T, B_0)= 8 \ln(2) (g/2)^2 \mu_B^2 \frac{M_\downarrow}{2\pi \hbar^2v_D^2} T
	\label{eq_muB0_chiT}
\end{align}
 Note that $g'$ has been set to be $2g$ to get this result. The linear $T$ behavior agrees with the data shown in Fig. 1c inset in the main text. By fitting the coefficient of the linear $T$ term we extract the parameter $v_D\sqrt{9/M_\downarrow}$ to be 5150 m/s =$34.1 meV$ $\cdot$\AA  \ by taking $g=2$, in excellent agreement with the value obtained from the linear slope in $|B-B_0|$ in the main text.

Next, to further explain the temperature dependence of $\chi$ in Fig. 1(c) in the main text, we plot the same data points in  Fig.~\ref{FigS_M}(b). We begin with a simple valley model, assuming a $V$-shaped magnetic susceptibility near the Dirac point $B_0$ as shown in Fig. \ref{FigS_thermalsmearing}(a), where $\chi(T,B)$ is linear for $|B-B_0|<B_p$ with a minimum at $B_0$, and $\chi=\chi_p$ otherwise. We can evaluate Eq. \ref{eq_muB_smear} to obtain an analytical solution of $\chi(T,B)$. A convolution integration between the two terms plotted in Fig. \ref{FigS_thermalsmearing}(a) gives:
\begin{equation}
	\chi(T,B)=\chi_p+\chi_p \frac{T}{T_p}\ln[\frac{(1+e^{\frac{-\frac{1}{2}g^\prime \mu_B (B-B_0)}{k_B T}})^2}{(1+e^{\frac{-\frac{1}{2}g^\prime \mu_B (B-B_0)}{k_B T}}e^{\frac{T_p}{T}})(1+e^{\frac{-\frac{1}{2}g^\prime \mu_B (B-B_0)}{k_B T}}e^{-\frac{T_p}{T}})}]
	\label{Diracpoint_smear}
\end{equation}
where $B_0$ is the magnetic field value at the Dirac point, $B_p$ is the half-width of the V-shape, and $T_p=\frac{\frac{1}{2}g'\mu_B B_p}{k_B}$. A plot of $\chi(B)$ at different temperatures is shown in Fig. \ref{FigS_thermalsmearing}(b), where the minimum in $\chi(B)$ at the Dirac point increases with temperature. Setting $B=B_0$, Eq. \ref{Diracpoint_smear} can be simplified to:
\begin{equation}
	\chi(T,B_0)=\chi_p[1+ \frac{T}{T_p}\ln(\frac{2}{1+\rm{cosh} (\frac{T_p}{T})})]
	\label{Dp_smear_B0}
\end{equation}
A plot of Eq. \ref{Dp_smear_B0} of $\chi(B_0)$ versus temperature is shown in Fig. \ref{FigS_thermalsmearing}(c). At the low-temperature limit, when $T\ll T_p$, Eq. \ref{Dp_smear_B0} goes to $\chi(T,B_0)\approx \ln (4) \chi_p \frac{T}{T_p}$. As evident in the plot, $\chi(B_0)$ is $T$-linear in low temperatures near the ground state, and gradually saturates as $T$ rises. Additionally, an offset $\chi_c$ should be added considering a paramagnetic contribution to $\chi(T,B)$:
\begin{equation}
	\chi(T,B_0)=\chi_c+\chi_p[1+ \frac{T}{T_p}\ln(\frac{2}{1+\rm{cosh} (\frac{T_p}{T})})]
	\label{Dp_smear_chi_c}
\end{equation}

In Fig. \ref{FigS_M}(b), the data are fitted well by an orange curve using Eq. \ref{Dp_smear_chi_c}, which indicates the 1/9 magnetization plateau could have a fermionic origin. The fitting parameters are $\chi_c=0.0019~\mu_B/\rm{T/Cu^{2+}}$, $\chi_p=0.0039~\mu_B/\rm{T/Cu^{2+}}$, and $T_p=\frac{\frac{1}{2}g^\prime \mu_B B_p}{k_B}=9.5$ K. If we use $g^\prime=4$, then $B_p=7.1$~T, which is consistent with the half-width of the valley in the inset of Fig. 1(c) in the main text. The black dashed linear line is the low-temperature limit ($T\ll T_p$) of Eq. \ref{Dp_smear_chi_c}, whose expression is $\chi (T)=\chi_c+\ln (4)\chi_p \frac{T}{T_p}$. Notably, at $T<3$~K, the data in Fig. \ref{FigS_M}(b) show a linear $T$-dependence with a positive slope. This $T$ linear behavior as $T\rightarrow 0$ is in sharp contrast to the gapped behavior that has been reported at the 1/3 plateau in other AF systems which can be understood in terms of magnon excitations \cite{Sheng2022, Kout2015}, and also differs from the incompletely developed plateau \cite{Smirnov2017}, as shown in Fig. \ref{FigS_M}(c).

\section{Model of the quantum oscillations of the field-driven Fermi surfaces of Dirac fermions}\label{sec_QO}

Here, we use a simple model to describe the quantum oscillations of the Dirac fermions in the U(1)-Dirac spin liquid state under the magnetic field. The energy dispersion of two-component Dirac fermions can be described by the equation $E(\textbf{k})=\hbar v_D k$, and $v_D$ is the Fermi velocity. After applying the magnetic field, the chemical potential $\mu$ will split as $\mu_{\downarrow(\uparrow)}= Sg^{\prime}\mu_B (\pm B-B_0)$ due to the Zeeman effect. Here $S=1/2$, and $g$-factor is around 2 in the YCOB. We assume the down spins will approach the Dirac point at $B=B_0$, so in the following derivation, we mainly focus on the energy of down spins. Therefore, we can set $\mu_{\downarrow}=\frac{1}{2}g^{\prime}\mu_B(B-B_0)=E_F$ to get the expression of Fermi wavelength $k_F$: 
\begin{align}
	k_F=\frac{g^{\prime}\mu_B}{2 \hbar v_D} (B-B_0)
	\label{eq_kF}
\end{align}
Next, we consider the Landau level quantization of Dirac fermions under the magnetic field. Assume spinors see gauge magnetic field $b=\alpha B \rm{cos}\theta$. When $\mu_B B>> \Delta E_L$ where $\Delta E_L$ is the energy of Landau level spacing, we can use Bohr-Sommerfeld semiclassical quantization \cite{Shoenberg}:
\begin{align}
	\frac{eb_n(n+\gamma_0)}{\hbar}=\frac{A_F}{2\pi}
	\label{eq_Onsager1}
\end{align}
Here $b_n$ is the gauge field when $n$th Landau level crosses Fermi energy, and $0\leq \gamma_0<1$ is the phase factor, which can be dependent on the direction of the magnetic field. $A_F=\pi k_F^2$ is the Fermi surface area. After plugging in $b_n$ and $k_F$, we can get 
\begin{align}
	\frac{(B_n - B_0)^2}{B_n}=\frac{8 e \alpha \hbar v_D^2}{g^{\prime 2} \mu_B^2}\rm{cos}\theta (n+\gamma_0)
	\label{eq_Bn_period}
\end{align}
Finally,  the period of quantum oscillations is 
\begin{align}
	\frac{(B_{n+1} - B_0)^2}{B_{n+1}}-\frac{(B_{n} - B_0)^2}{B_{n}}=\Delta B 
	\label{eq_F_period}
\end{align}
where we set $\Delta B=\frac{8 e \alpha \hbar v_D^2}{g^{\prime 2}\mu_B^2}\textup{cos}\theta$ which is Eq. 4 in the text.
Therefore, we find that the oscillations of Dirac fermions are periodic in $(B-B_0)^2/B$, which is very different from the usual $1/B$ period behavior in the metals and topological semimetals. Using value $\Delta B/\textup{cos}\theta=3.35$ T obtained from fitting in Fig. 5(c) in the main text, we can estimate the $\sqrt{\alpha}$ times the Dirac velocity as done in the text.

\section{ Expression for the torque}\label{sec_tor}

Here we derive a simplified expression of the torque that accounts for the quantum oscillations. For a single Dirac node with velocity $v_D$ and chemical potential $\mu$, the free energy per area takes the form \cite{Thesis,Igor2011}
\begin{equation}
	\frac{F}{L^2}=-\frac{2\omega_D^2}{\pi v_D^2}\frac{T}{\sinh{(\frac{4\pi^2T\mu}{\omega_D^2})}} \cos{(\frac{2\pi \mu^2}{\omega_D^2}-\varphi)}
	\label{eq_F}
\end{equation}
where $\omega_D^2=v_D^2(2e/\hbar)|B_c|$ and $B_c$ is the component of $B$ along the c-axis. The phase $\varphi$ is related to  $\gamma_0$ introduced in the last section by $\varphi=2\pi (\gamma_0-1/2)$. For conventional quadratic bands,  $\varphi=0$ corresponds to $\gamma_0=1/2$, which reflects the factor 1/2 in Landau levels that goes as $n+1/2$. For the Dirac spectrum in graphene,  $\varphi=\pi$.  The difference has to do with the Berry phase in graphene. Since we do not have a microscopic understanding of the origin of the band, we leave $\varphi$ as a parameter that is independent of B or the chemical potential.   To apply this to the current problem, we replace $B_c$ by $\alpha B_c=\alpha B \cos{\theta}$ and $\mu$ by $\mu_\downarrow = (g'/2)\mu_B(B-B_0)$. Note that the magnetic field plays a dual role: its c component is responsible for the orbital motion and quantization while its magnitude is responsible for the chemical potential, hence  $F(B,B_c)$. 
Defining $B_a$ as the field along the $a$ axis, we want to calculate $M_a=\partial F/\partial B_a$ and $M_c=\partial F/\partial B_c$. To do this, we need to include the implicit dependence on $B_a$ and $B_c$ via $B=\sqrt{B_a^2+B_c^2}$. This creates a rather complicated formula. Fortunately, we can show that for the torque, we need to take the derivative with only the explicit dependence on $B_c$.  The torque is given by $\vec{\tau}=\vec{M} \times \vec{B}$, and  its component along $b$ is denoted by $\tau=M_cB_a-M_aB_c$ where $M_a=\partial F/\partial B_a 
|_{B_c}$ and $M_c=\partial F/\partial B_c |_{B_a}$. Since $F$ is a function of $B$ and $B_c$, we need to write $B=\sqrt{B_a^2+B_c^2}$ and use the implicit dependence on $B_a$ and $B_c$. This gives rise to  $M_a=(B_a/B)\partial F/\partial B $ while $M_c=(B_c/B)\partial F/\partial B  + \partial F/\partial B_c |_{B}$. Putting these in the expression for $\tau$ we find that the term involving $\partial F/\partial B$ cancels and we are left with
\begin{equation}
	\tau/B= \frac{B_a}{B}(\partial F/\partial B_c)|_B.
	\label{eq_toroverB}
\end{equation}
From Eq. \ref{eq_F}, we see that $F$ depends on $B_c$ both in the prefactor and inside the cosine term. We will now argue that the derivative of the cosine term dominates. Let us re-write Eq. \ref{eq_F} as  $F \propto B_c^2[\frac{aT}{\mathrm{sinh}(aT)}]\mathrm{cos}X$ where $a=2\pi X/\mu$ . At low $T$ so that $aT<<1$, the LK factor $[\frac{aT}{\mathrm{sinh}(aT)}]$ goes to unity. Using $X \propto 1/B_c$ we find $\partial 
F/\partial B_c \propto B_c(2\mathrm{cos}(X)+X\mathrm{sin}(X))$  
Since $X=2\pi n$ for the nth oscillation, the second term, which is proportional to $X$, dominates as long as $n>1$. In other words, there is a correction to the phase factor $\varphi$ equal to $2/X \approx 1/(\pi n) $, which we will ignore. For $aT>>1$, the LK factor $[\frac{aT}{\mathrm{sinh}(aT)}]$ goes as $exp(-aT)$. $\partial F/\partial B_c \propto B_c(aT\mathrm{cos}(X)+X\mathrm{sin}(X))$. Since $aT/X=2\pi T/\mu$, the first term can be ignored if $T<<\mu$. Therefore we keep only the term coming from differentiating the cosine term in Eq. \ref{eq_F}, resulting in Eq. 3 in the main text.

\section{Procedure to obtain the period $\Delta B$}\label{sec_DeltaB}

Here we describe the procedure to obtain the period $\Delta B$. First, we consider the spinon-chargon mechanism for the origin of gauge field $b$ where $b=\alpha B$ and $\alpha$ is a constant. We will be fitting either the first or the second derivative of $\tau$. Focusing on the oscillatory term and keeping track of the overall sign, the first derivative of Eq. 3 in the main text simply converts the sine to a cosine. 
\begin{equation}
	\frac{d\tau}{dB} \propto   
	-\text{sin}(\theta) \textup{cos}(2\pi \frac{(B-B_0)^2}{\Delta B \cdot B}-\varphi)
	\label{eq_derivative1}
\end{equation}

For $\theta>0$, we assign the maximum value of $d\tau/dB$ in Fig. 5(a) in the main text as integer $n$ and the minimum value as half-integer, then we get the following Landau index equation:
\begin{align}
	2\pi n+\pi &= 2\pi \frac{(B-B_0)^2}{\Delta B \cdot B}-\varphi  \nonumber \\ 
	n &= \frac{(B-B_0)^2}{\Delta B \cdot B} - \frac{\varphi+\pi}{2\pi}
	\label{eq_Landauindex_1st}
\end{align}
The fitting results are shown in Fig. 5(b) and (c) in the main text.

For the second derivative, it is convenient to re-write the resulting sine in terms of cosine by introducing a $\pi /2$ phase shift:
\begin{equation}
	\frac{d^2\tau}{dB^2} \propto  \text{sin}(\theta) \textup{cos}(2\pi \frac{(B-B_0)^2}{\Delta B \cdot B}-\varphi^\prime)
	\label{eq_derivative2}
\end{equation}
where  $\varphi^\prime=\varphi+\pi /2$. The definition of $\Delta B$ is given by Eq. 4 in the main text. Similarly, here we assign the maximum value of $d^2\tau/dB^2$ 
as integer $n$ and the minimum value as half-integer, then the Landau index equation for the second derivative is:
\begin{align}
	n &= \frac{(B-B_0)^2}{\Delta B \cdot B} - \frac{\varphi^\prime}{2\pi}
	\label{eq_Landauindex}
\end{align}

Note that for $\theta$ negative, the even and odd integer assignment is reversed because of the overall minus sign.  We have fitted the second derivative data as shown in Fig. \ref{FigS_Landau} using Eq. \ref{eq_Landauindex}. 
The fitting results are shown in Fig. 5(c) in the main text. To get the correct positions of extremal points in Fig. \ref{FigS_Landau} bottom panels, the oscillation patterns need to be filtered out by subtracting a Loess-smoothed background from the $d^2\tau/dH^2$ data as shown in Fig. \ref{FigS_Landau} top panels.

Next, we discuss our criteria for selecting effective oscillation field windows at different angles. We present data for twelve angles in Fig. \ref{FigS_Landau} to illustrate all the angles ranging from -32.6$^{\circ}$ to 84.6$^{\circ}$. At the low-angle region ($|\theta|<50^{\circ}$), the oscillation patterns are very similar except for the shorter periods with increasing angles, so we can build consistent criteria here with six extrema included when $|\theta|<20^{\circ}$ or seven extrema when $|\theta|>20^{\circ}$ in Fig. \ref{FigS_Landau}(a-f). When $|\theta|>50^{\circ}$, such as 62.8$^{\circ}$ in Fig. \ref{FigS_Landau}(f), the oscillation period is shorter and more oscillations appear, so we include every extremum up to the point when data become noisy. This happens for angles larger than 50$^{\circ}$, and we keep data up to 33~T. The exception is that at the largest angle  84.6$^{\circ}$, there is a sharp dip around 31 T, and beyond this, the features show a strong temperature dependence and no longer resemble oscillations (see Fig.\ref{FigS_twoTdepend}). As yet, we do not understand the origin of this structure, so we did not include this dip and its associated features in our oscillation analysis.

Once we have labeled the Landau indices, we apply Eq. \ref{eq_Landauindex} to fit the data. There are 3 variable parameters in the fitting: $B_0, \Delta B$, and $\varphi'$. $B_0$ corresponds to the magnetic field where the Dirac point is located, and $\Delta B$ is the parameter representing the interval between oscillation extrema, in analogy to the period of magnetic oscillations. The uncertainty of $B_0$ and $\Delta B$ contains two contributions. The first is the uncertainty when picking the positions of extrema. We estimated the field position uncertainty of extrema to be $\sigma_B=0.5$ T [see Fig. \ref{FigS_Landau}]. Using the 
law of propagation of uncertainties, we estimated the uncertainty of $B_0$ to be $\sigma_{B_0}=\frac{\sigma_B}{B_{ave}}B_0$, where $B_{ave}\approx30$ T is the average magnetic field of the oscillation region. Similarly, the uncertainty of $\Delta B$ is $\sigma_{\Delta B}=\frac{\sigma_B}{B_{ave}}\Delta B$. Another uncertainty comes from the fitting error of Eq. \ref{eq_Landauindex}. These two uncertainties are added together and displayed in Fig. 5(c) in the main text as error bars. We choose $\varphi'$ to be constant independent of angle. We perform fitting for the entire data set of angles using a fixed $\varphi'$ and, in the end, find the optimal $\varphi'$ which minimizes the error: $\varphi'/2\pi = -0.45$, ie, $\varphi /2\pi = -0.7$. 
However, the minimum for $\varphi'$ is relatively shallow, which indicates that our value of $\varphi$ is slightly uncertain; hence, we do not discuss its significance further here. Nevertheless, to put it in perspective, in a conventional metal with a parabolic band, $\varphi=0$, whereas for the Dirac spectrum in graphene, $\varphi /2\pi=0.5$.

Next, we discuss fitting the the period using the DM mechanism as the origin of $b$. In this 
case,  $b=\gamma  \langle \vec{S_z} \rangle $, and we can rewrite it as $b= \gamma^{\prime}  M(B)\rm{cos}\theta$ if we ignore the $g$-factor anisotropy as we did in deriving Eq. 3 in the main text. If $M$
is strictly proportional to $B$ and given by $\chi_{ave} B$, where $\chi_{ave}$ is the average spin susceptibility, this model will be identical to the spinon-chargon model. In practice, near the $\frac{1}{9}$ plateau, the magnetization is not linear in $B$. It is convenient for us to explicitly parametrize this deviation from linearity and cast the result in a form as similar to the spinon-chargon case as possible. To this end, we introduce the dimensionless ratio $R(B)=M/(\chi_{ave} B)$. We will see that this ratio is close to unity. We can write $b=R(B)
\gamma^\prime \chi_{ave}   B\cos{\theta} $. We can again derive Eq. 4, except that now 
\begin{equation}
	\Delta B =R(B) B_{DM} \text{cos}(\theta)
	\label{eq_DB_DM}
\end{equation}
where $B_{DM}$ is a constant. We compute $R(B)$ from our magnetization measurements and use Eq. \ref{eq_DB_DM} to fit the data to extract $B_{DM}$ as a function of $\theta$. To facilitate comparison with the spinon-chargon mechanism, we introduce $\widetilde{\Delta B}=B_{DM} \text{cos}(\theta)$ and check how well it agrees with $\text{cos}(\theta)$ as a test of the DM mechanism.

We conduct an 8th-order polynomial fit to the plot of $M(B)$ versus $B$ as shown in Fig. \ref{FigS_DMmodelfit}(a).  In other words, $M_(B) \approx \sum_{i=0}^{8}A_i\cdot \textit{B}^i$, where $A_i$ are the fitting constants obtained from 8th-order polynomial fitting. In the inset we show the ratio $R(B)$. For the $B$ field range of interest between 20 T and 40 T, the deviation from unity is less than 12\%.
Instead of Eq.\ref{eq_Landauindex}, we make use of Eq. \ref{eq_DB_DM} to arrive at the following equation to fit the Landau index,
\begin{equation}
	n = \frac{(B-B_0)^2}{\widetilde{\Delta B}\cdot \sum_{i=0}^{8}A_i\cdot \textit{B}^i /\chi_{ave}}-\frac{\varphi^{\prime}}{2\pi}
	\label{eq_DMfit}
\end{equation}
Applying the equation above to fit the data points in Fig. \ref{FigS_Landau}, we can get the angular dependence of $\widetilde{\Delta B}$ and $B_0$, as shown in Fig. \ref{FigS_DMmodelfit}. The fitted value of  $\widetilde{\Delta B}$ at 0$^{\circ}$ is 3.35 T, which is the same as the  $\Delta B$ at 0$^{\circ}$ in Fig. 5(c) in the main text.

\section{FFT analysis of magnetic oscillations}\label{sec_FFT}
	
In order to conduct FFT analysis on the oscillatory features, we subtract a background from the derivative data; the background is based on the Loess-smooth method as shown in Fig. \ref{FigS_toranalysis}(a). The resulting background-subtracted temperature dependence of the magnetic oscillations is shown in Fig. \ref{FigS_toranalysis}(b) top panel. Next, we rescaled the $x$ axis of Fig. \ref{FigS_toranalysis}(b) from $B$ to $(B-B_0)^2/B$ in order to get periodic patterns along the $x$ axis. Finally, we conducted standard FFT analysis on the periodic patterns, and the corresponding FFT spectra are shown in the inset of Fig. 3(c) in the main text at different temperatures using a transform window from 24.0~T to 35.7~T. We can see there is a dominant frequency around 0.35~T$^{-1}$, which corresponds to our defined period $\Delta B \approx 2.86$~T at -32.6$^{\circ}$. The amplitudes of the dominant frequency at different temperatures are plotted in Fig. 3(c). We also notice that the frequency becomes slightly smaller with increasing temperature; one possible reason is the linear-energy-dispersion-induced frequency shift at different temperatures \cite{Guo2021}. We emphasize that the mass obtained by LK fitting $aT/\mathrm{sinh}(aT)$ in Fig. 3(c) in the main text is the averaged effective mass over the FFT window; in fact, $m^*$ is field-dependent. Therefore, we also show the amplitude of each peak and valley from the top panel of Fig. \ref{FigS_toranalysis}(b) at fixed fields in Fig. \ref{FigS_toranalysis}(c), and find $m^*$ by performing LK fits at each field. The consequent field dependence of the mass is shown in the inset of (c). The increasing trend of $m^*$ with magnetic field is roughly consistent with Eq. 5 in the main text. 

\section{Magnetization plateaus and oscillations in different samples and a control sample}\label{sec_diffsample}

\textcolor{black}{
	To verify the repeatability of oscillations in different samples and different magnets, we conducted the magnetic torque measurements in YCOB sample S5 in the Mid-Pulse magnet. The temperature dependence of torque and torque second derivative are shown in Fig. \ref{FigS_pulsetor}, the torque and $M_t$ data in Fig. \ref{FigS_pulsetor}(a) are very similar to the data in Fig. 3(a) in the main text except for the field limit in Fig. \ref{FigS_pulsetor}(a) reach up to 59 T which is in $\frac{1}{3}$ magnetization plateau region. More details can be seen in the torque second derivative in Fig. \ref{FigS_pulsetor}(b), which also has similar oscillation patterns roughly periodic in $B$ compared with Fig. 3(b), but with more sets of QOs showing up at the higher field, and the oscillation amplitudes will decay quickly if the temperature increase. }

\textcolor{black}{A control experiment is further carried out to demonstrate that these exciting properties have an intrinsic origin. We grew a control sample with the  YCOB-58\%Cl, which has 58\% Br atoms replaced by Cl (See \textcolor{blue}{section Materials and Methods} in the main text for sample detail). According to Ref. \cite{Xu2023}, different from the QSL ground state proposed in YCOB, YCOB-58\%Cl has a long-range order forming at 7 K, and the ground state of YCOB-58\%Cl is an AF-ordered state, even though the crystal structures are identical to the undoped YCOB. The comparison is shown in Fig. \ref{FigS_Cl0.5compare}.  The torque second derivative of both samples at 0.5 K and -12.6 $^\circ$ is shown in Fig. \ref{FigS_Cl0.5compare}, and the torque data are shown in the inset. In contrast to both $\frac{1}{9}$ and $\frac{1}{3}$ plateaus observed in undoped YCOB S5, the control sample does not reveal any of these features. Furthermore, very different from the abundant oscillation features in YCOB, the torque second derivative of the control sample YCOB-58\%Cl is flat and featureless except for noise, and the corresponding torque data is a well-defined quadratic curve at high field which is consistent with the constant susceptibility behavior. These very different features in undoped YCOB and the control sample YCOB-58\%Cl strongly indicate the intrinsic properties of the observations of the plateaus and magnetic oscillations. }

Overall, magnetic torque measurements were conducted in YCOB sample S1, S2, S5 and YCOB-D sample D1, D2 in three different magnets with two different methods (See \textcolor{blue}{section Materials and Methods} in the main text). The corresponding figures are shown in Fig. \ref{FigS_compare}. Black dashed lines marked five extrema values corresponding to the magnetic oscillations. The oscillation patterns align well, demonstrating the robust and intrinsic nature of these features. However, there are significant differences between these samples for features below 20~T. While sample S1 and S2 are still in agreement, the deuterated samples look different and are also different from each other. Note that 20~T is near or below the plateau magnetic field $B_0$.  The data below this field seems to exhibit two large peaks, which are sample dependent. In view of these uncertainties, we do not attempt to analyze these peaks in this paper and instead focus on the structures for $B>B_0$ which are highly reproducible.

\section{ Quantum oscillation from the spin up Fermi surface}\label{sec_spinup}

Here, we comment on the possibility of seeing QO when the spin-up chemical potential lies near the top of band 4 and/or near the bottom of band 5. This happens near $B=B_0$ when band 4 and 5 overlap or when the gap between them is small. Surprisingly we find that the quantum oscillation takes the same form as given in Eq. 3 in the main text, even though they are from a quadratically dispersing band. Let us consider the case where $\mu_\uparrow$ lies just below the top of band 4. We can treat this as a hole band with quadratic dispersion with mass $m_h$. The QO is given by the standard formula
\begin{equation}
	M \propto [\frac{aT}{\mathrm{sinh}(aT)}]\mathrm{sin} (2\pi \mu_\uparrow/\omega_c)
	\label{eq_QO_upspin}
\end{equation}
where $\omega_c=e|\alpha| B_c/m_h$ and $a=2\pi^2/\omega_c$. What is interesting is the way $\mu_\uparrow$ depends on $B-B_0$. We have seen earlier that it is almost pinned because the DOS for the up spin is much larger than that of the down spin which is near the Dirac node. In Eq. \ref{eq_mu_uparrow2}, we show that it goes as $(B-B_0)^2$. Putting Eq. \ref{eq_mu_uparrow2} into \ref{eq_QO_upspin}, we find that the oscillatory part takes the same form as Eq. 3 in the main text. The LK factor now takes the standard form, and the mass extracted is given by $m^*=m_h/|\alpha|$. Thus, quantum oscillations from the up-spin Fermi surface may show up as well, with a different $\Delta B$ from that from the down-spin Dirac fermions. 

\section{ Estimate of Fermi velocity based on magnetic susceptibility}\label{sec_velocity}

Another method to estimate the Fermi velocity is to investigate the magnetic susceptibility data in Fig. 1(c) in the main text. Start from Eq. \ref{eq_muB_chi}, when $H_{\parallel}$ field is around 20 T, we know $\mu_{\downarrow}$ is located at Dirac point, which means $D(\mu_{\downarrow})\ll 1$. Therefore, Eq. \ref{eq_muB_chi} can be simplified to $\chi \simeq 2 g \mu_B^2 D(\mu_{\downarrow})$. We know for a 2D Dirac system, the density of states per unit area is
\begin{align}
	D(\mu)=\frac{M_{\downarrow} |\mu|}{2\pi \hbar^2 v_D^2}
\end{align}
Here $M_{\downarrow}$ is the number of Dirac nodes in the original Brillouin zone. For down spins, we can plug in $\mu_{\downarrow}=\frac{1}{2}g^{\prime}\mu_B(B-B_0)$. After taking $g^{\prime}=2g$, we can get 
\begin{align}
	\chi \simeq \frac{M_{\downarrow} g^2 \mu_B^3 (|B-B_0|)}{\pi \hbar^2 v_D^2}
\end{align}

This means that around Dirac point $B_0$, $\chi$ is linear with a magnetic field. Fig. 1(c) in the main text verified this linear relation. Now we can use the linear slope around $B_0$ to estimate $v_D$. Define the slope of the linear region in $dM/dH$ versus $H$ around Dirac point $B_0$ as $k_{\chi}$ in Fig. 1(c) when $B \parallel c$. Therefore, if we consider the slight asymmetry of the Dirac cone, we can find the slope of the left side from 15 T to 19 T is $k_{\chi-L}$ = -87.3 J T$^{-3}$ m$^{-3}$, while the slope of the right side from 19 T to 25 T is $k_{\chi-R}$ = -57.2 J T$^{-3}$ m$^{-3}$ by estimating 0.6 K magnetic susceptibility data. The expression of slope $k_{\chi}$ is
\begin{align}
	|k_{\chi}|= \frac{M_{\downarrow} g^2 \mu_B^3}{\pi \hbar^2 v_D^2} \cdot \frac{1}{c}
\end{align}
Here $c=5.99\ \textup{\AA}$ \cite{Zeng2022} is the lattice constant in $c$ axis, which is included to convert the unit of $\chi$ from per unit area to per unit volume in order to compare with experiments. We estimate $v_D=3962 (4898) \sqrt{M_{\downarrow}/9} $ m/s  which equals $ 26.2 (32.4)  \sqrt{M_{\downarrow}/9}~$meV$\cdot$\AA ~for the left (right) side of the Dirac cone. Note that this estimate is sensitive to the number of Dirac nodes per Brillouin zone $M_{\downarrow}$, which is not known. However, due to the tripling of the unit cell, $M_{\downarrow}$ comes in multiples of 3. If the Dirac node is at the $\Gamma$ point, $M_{\downarrow}=3$. If not, the next value is $M_{\downarrow}=9$ due to 3-fold symmetry.  For $M_{\downarrow}=9$, we find $v_{\rm D} \approx 26.2 (32.4)~$meV$\cdot$\AA ~for the left (right) side of the Dirac cone.

\section{ Estimate of the ratio $\alpha$ between the gauge magnetic field and applied magnetic field in the spinon-chargon mechanism.}\label{sec_gaugefield}

We discuss the value of $\alpha$, which is defined as the ratio between the gauge magnetic field $b$ and the $c$ component of the physical magnetic field $B \cos(\theta)$ in the spinon-chargon mechanism. In part of the treatment for the Hubbard model, the electron is decomposed into fermionic spinon and relativistic bosonic chargon. \cite{LeeSS} The constraint between spinon and chargon leads to a coupling of the spinon to the external field even though it is charge neutral. The physics is captured by the Ioffe-Larkin relation and has been recently discussed and slightly extended \cite{Lee06,Inti22}. The result is that
\begin{equation}
	\alpha=b/B \cos(\theta)= \chi_o/(\chi_o+\chi_f+\chi_a) 
	\label{eq_Ioffe_Larkin}
\end{equation}
where $\chi_o,\chi_f$ are the orbital diamagnetic susceptibility of the chargon and spinon, and $\chi_a$ is from the Maxwell term for the gauge field due to integrating out higher energy degrees of freedom. In the free fermion model, $\chi_f=1/(12\pi m_f)$ which is of order $J$ for a spinon fermi surface with mass $m_f$ and smaller for a Dirac fermion. We expect $\chi_a \approx J $. On the other hand, $\chi_o \approx v_D^2/\Delta_c$ where $\Delta_c$ is the charge gap which is less than $U$ \cite{Dai}.
The bosons hop with amplitude of order $t$, so we expect $\chi_o > t^2/U \approx J$.
Therefore $\alpha$ can be close to unity.  However, we should add that this discussion is in the domain of low-energy effective theory, and cannot be carried over to the  case 
$U/t>>1$ where the charge is strongly localized. In that case, it is better to perform a $t/U$ expansion, following Ref. 
~\cite{Motrunich}. There it was found that there is a linear coupling between the scalar chirality $C$ and the physical magnetic flux through a triangle. The coefficient of that coupling was found to be $t^3/U^2$. On the other hand, the restoring force for the emergent gauge field $b$ is expected to be of order $J=t^2/U$ from integrating out high energy degrees of freedom. We expect $\alpha$ to be given by the ratio of these two coefficient and become of order $t/U$.

\newpage

\begin{table}[ht]
	\centering
	\begin{tabular}{|l|c|c|c|c|c|c}
		\hline
		\textbf{Sample name} & \textbf{ Magnet used} & \textbf{Physical property} & \textbf{$\frac{1}{9}$ plateau} & \textbf{High-field Oscillations} \\
		\hline
		YCOB M1  & Cell 1 in Los Alamos (65 T) & Magnetization & Yes & No \\
		& Duplex in Los Alamos (73 T) & Magnetization & Yes & No \\
		\hline
		YCOB M2  & Cell 3 in Los Alamos (65 T) & Magnetization & Yes & No \\
		\hline
		YCOB S1  & Cell 6 in Tallahassee (41 T) & Torque & Yes & Yes \\
		\hline
		YCOB S2  & Hybrid in Tallahassee (45 T) & Torque & Yes & Yes \\
		\hline
		YCOB-D D1 & Cell 6 in Tallahassee (41 T) & Torque & Yes & Yes \\
		\hline
		YCOB-D D2 & Cell 6 in Tallahassee (41 T) & Torque & Yes & Yes \\
		\hline
		YCOB S5 & Mid-pulse in Los Alamos (60 T) & Torque & Yes & Yes \\
		\hline
		YCOB-58\%Cl & Mid-pulse in Los Alamos (60 T) & Torque & No & No \\
		\hline
	\end{tabular}
	\caption{Summary of sample names, magnets used, physical quantities measured, presence of the 1/9 magnetization plateau, and observation of high-field oscillations following the 1/9 plateau for different samples. The comparison of all the torque curves is in Section \ref{sec_diffsample}, Fig. \ref{FigS_compare} and Fig. \ref{FigS_pulsetor}.}
	\label{tab_summary}
\end{table}

\newpage

\begin{figure}[!htb]
	\centering
	\includegraphics[width=1\columnwidth]{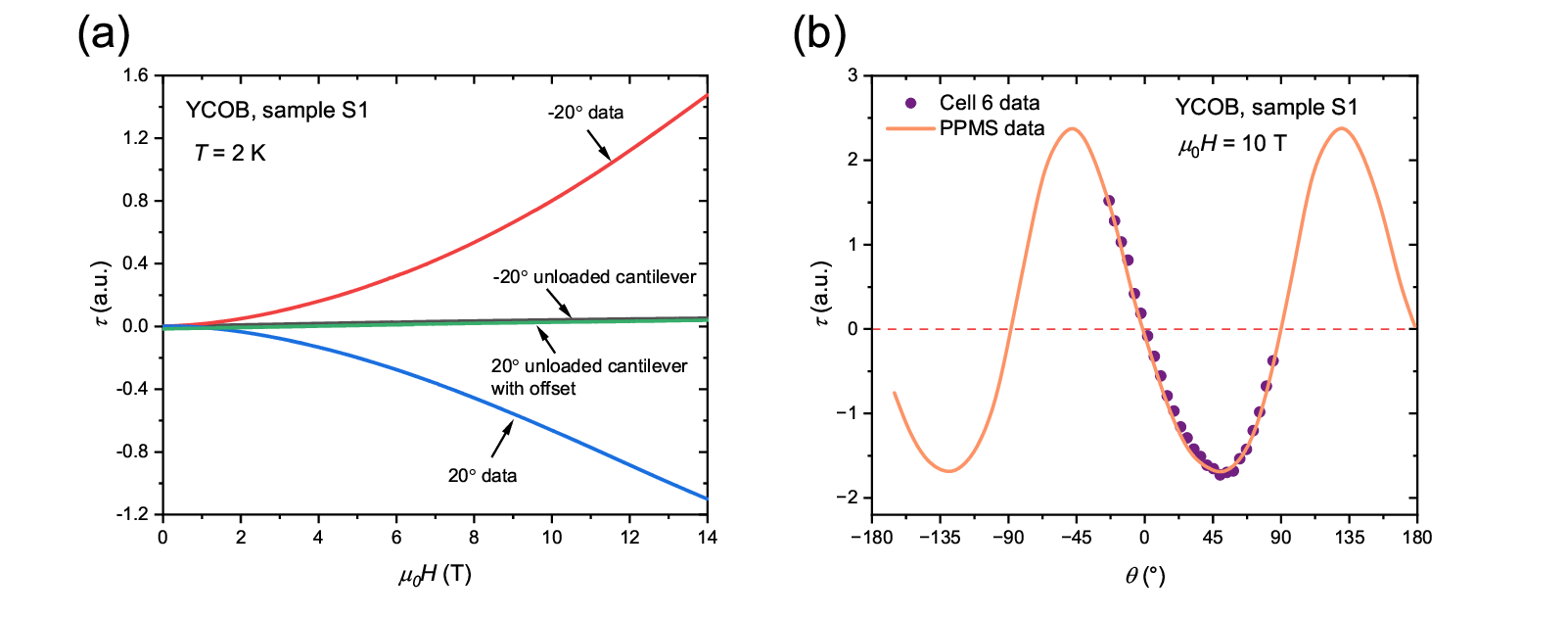}
	\caption{{\bf Background signal of cantilever setup and angle calibration.}
		(\textbf{a}) The torque signals of an empty cantilever versus $H$ field compared with the same cantilever setup loaded with YCOB sample S1  measured in PPMS at different angles. The empty cantilever signals are small enough and do not have angular dependence compared with the cantilever loaded with sample. (\textbf{b}) The angular dependence of torque signal comparison between PPMS and Cell 6 measurements of YCOB sample S1 at 10 T. The angles are shifted in Cell 6 measurements to match the PPMS data. This is the method to determine the angles in Fig. 2 and Fig. 5 in the main text. The torque signals have larger positive peaks, which indicates the piezoresistive cantilever setup does not have a perfectly symmetric response to tension and compression. 
	}
	
	\label{FigS_background}
\end{figure}

\begin{figure}[!htb]
	\centering
	\includegraphics[width=\columnwidth]{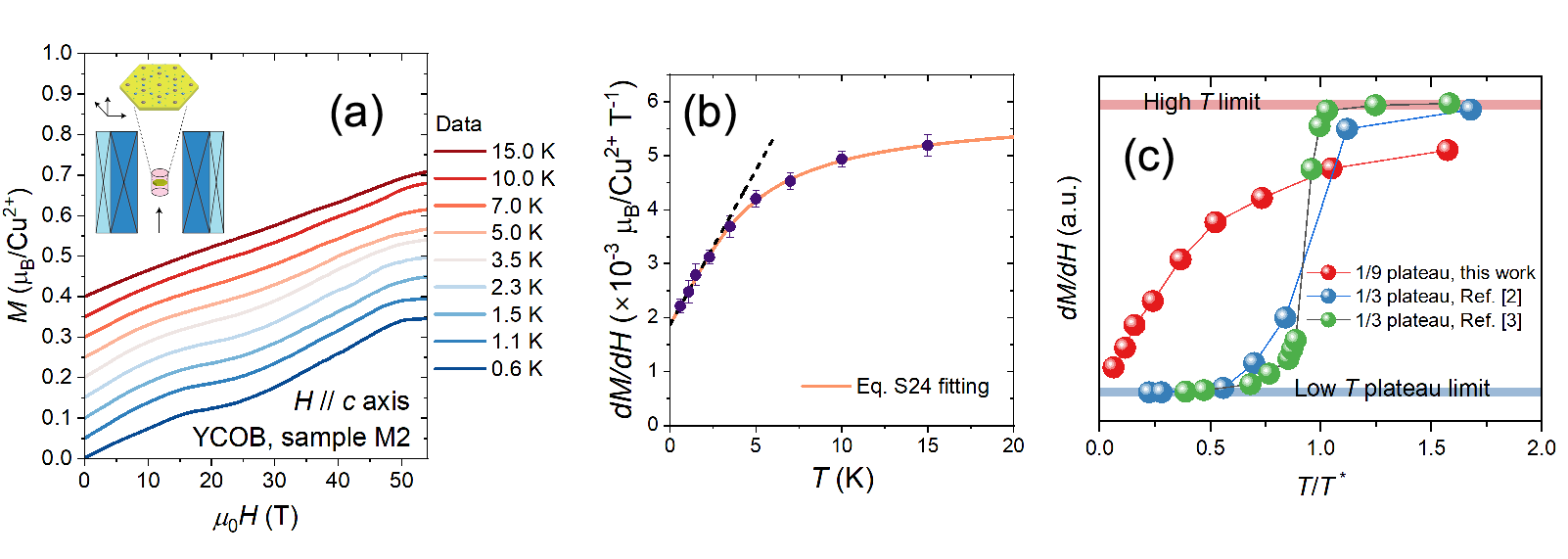}
	\caption{{\bf Temperature dependence of 1/9 magnetization plateau amplitude.}
		(\textbf{a}) Experimental setup of a compensated coil magnetometer for magnetization measurements in pulse magnetic field. The compensated coils are shown in dark blue (1000 turns) and light blue (500 turns in the other direction). The samples were stacked in a non-magnetic tube and inserted into the coil. 
		(\textbf{b}) The temperature dependence of the minimum value of $dM/dH$ in the $\frac{1}{9}$ plateau in the inset of Fig. 1(c) in the main text. The orange curve shows a fit by Eq. \ref{Dp_smear_chi_c} coming from the Fermi-Dirac thermal broadening model as described in Section \ref{sec_2gsmear}. The fitting parameters are $\chi_c=0.0019~\mu_B$/T/Cu$^{2+}$, $\chi_p=0.0039~\mu_B$/T/Cu$^{2+}$, and $T_p=~$9.5 K. The black dashed line is the low-temperature ($T\ll T_p$) linear approximation. 
		(\textbf{c}) The red dots are the data in (b) which is compared with that of the $\frac{1}{3}$ plateau states in other spin-$\frac{1}{2}$ triangular lattice quantum magnets Na$_2$BaCo(PO$_4$)$_2$ (blue dots) and Ba$_3$CoSb$_2$O$_9$ (green dots). For comparison, in YCOB, the temperature is scaled with $T^*=T_p=9.5$~K, and in the triangular lattice compounds, $T^*=T_N$ is the ordered temperature. 
	}
	\label{FigS_M}
\end{figure}

\begin{figure}[!htb]
	\centering
	\includegraphics[width=1\columnwidth]{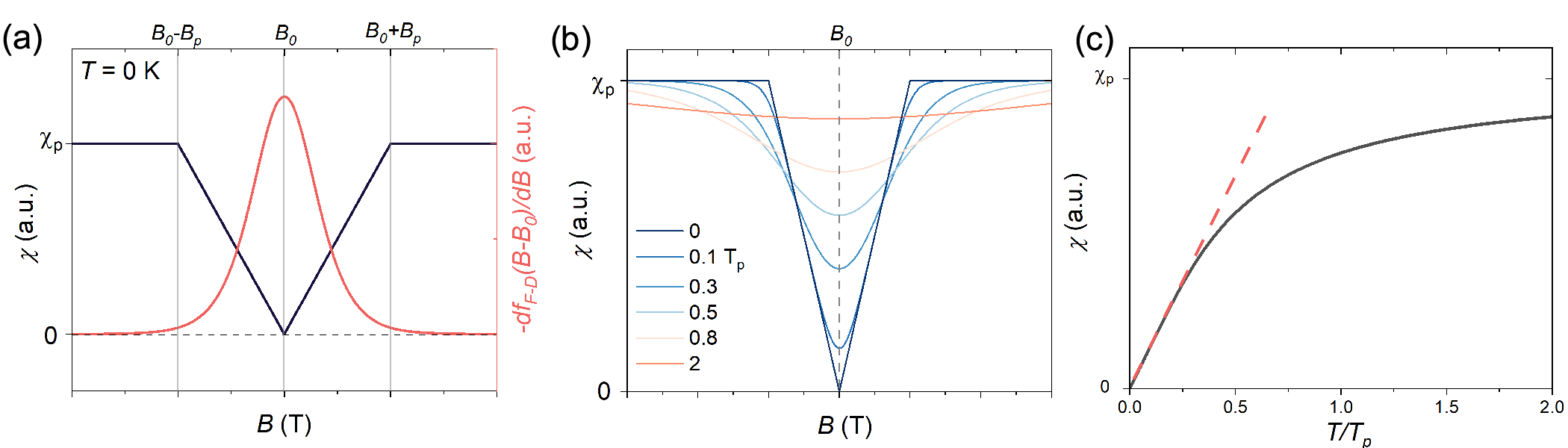}
	\caption{{\bf Thermal smearing model of the magnetic susceptibility at the Dirac point.}
		(\textbf{a}) Simulation of a V-shaped Magnetic susceptibility at zero temperature and $\frac{\partial f_{\rm{F}-\rm{D}}(B-B_0)}{\partial B}$ versus magnetic field. $B_0$ is the magnetic field at the Dirac point, and $B_p$ is the half-width of the V-shape. The dashed line marks the zero of the y-axis. (\textbf{b}) Simulation of thermal smearing of magnetic susceptibility $\chi(B)$ versus magnetic field.  (\textbf{c}) The simulated magnetic susceptibility at $B = B_0$ as a function of temperature. $\chi(B_0)$ is $T$-linear near the ground state. 
	}
	
	\label{FigS_thermalsmearing}
\end{figure}

\begin{figure}[!htb]
	\centering
	\includegraphics[width=0.7\columnwidth]{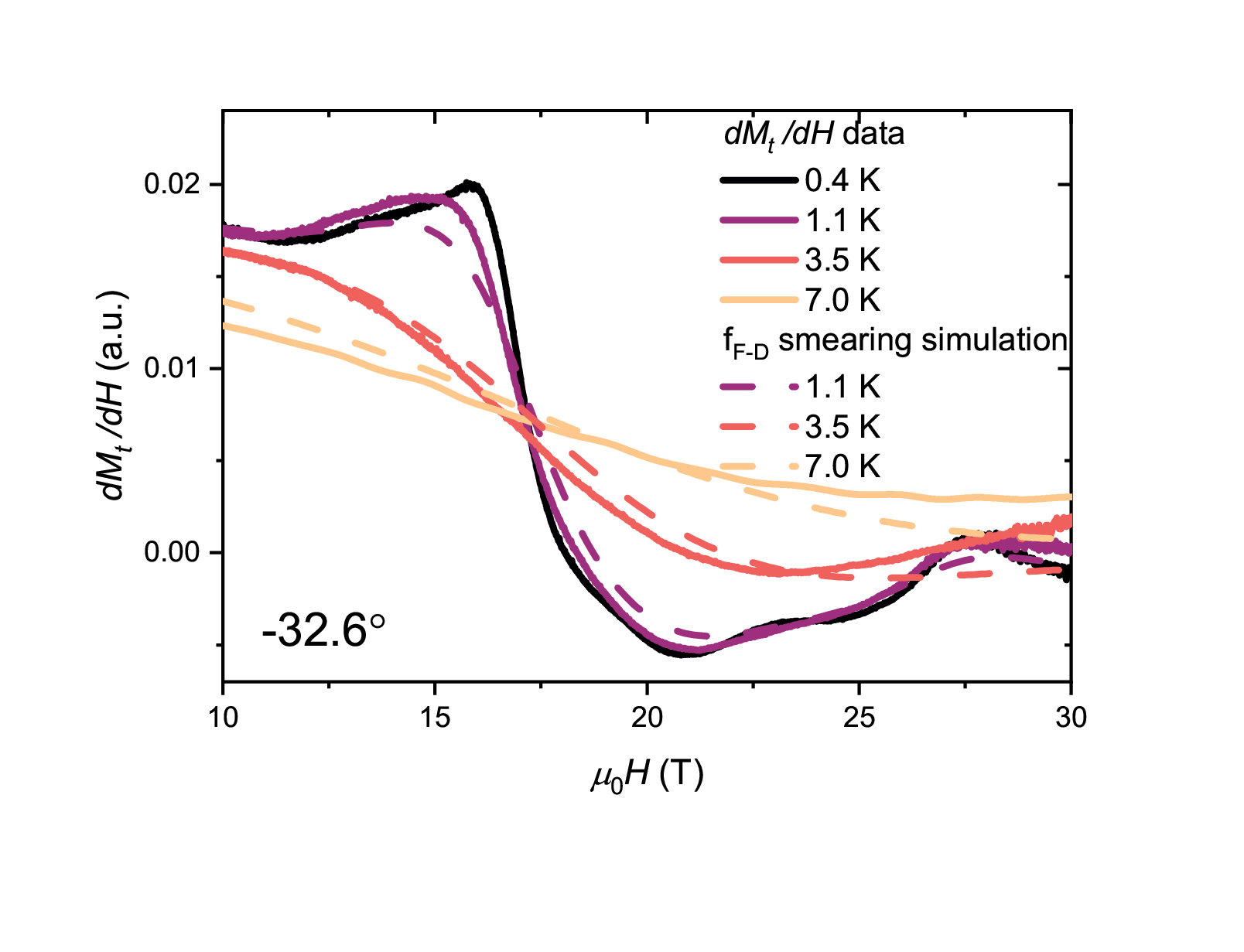}
	\caption{{\bf Thermal smearing model applied on the torque data near the Dirac point.}
		temperature dependence of ${\rm d}M_t/{\rm d}H$ data near $B_0$, and the calculated thermally smeared dash curves based on Fermi Dirac distributions $f_{F-D}$ (section \ref{sec_2gsmear}, Eq. \ref{eq_muB_smear}).
	}
	
	\label{FigS_tor_smear}
\end{figure}

\begin{figure}[!htb]
	\centering
	\includegraphics[width=1\columnwidth]{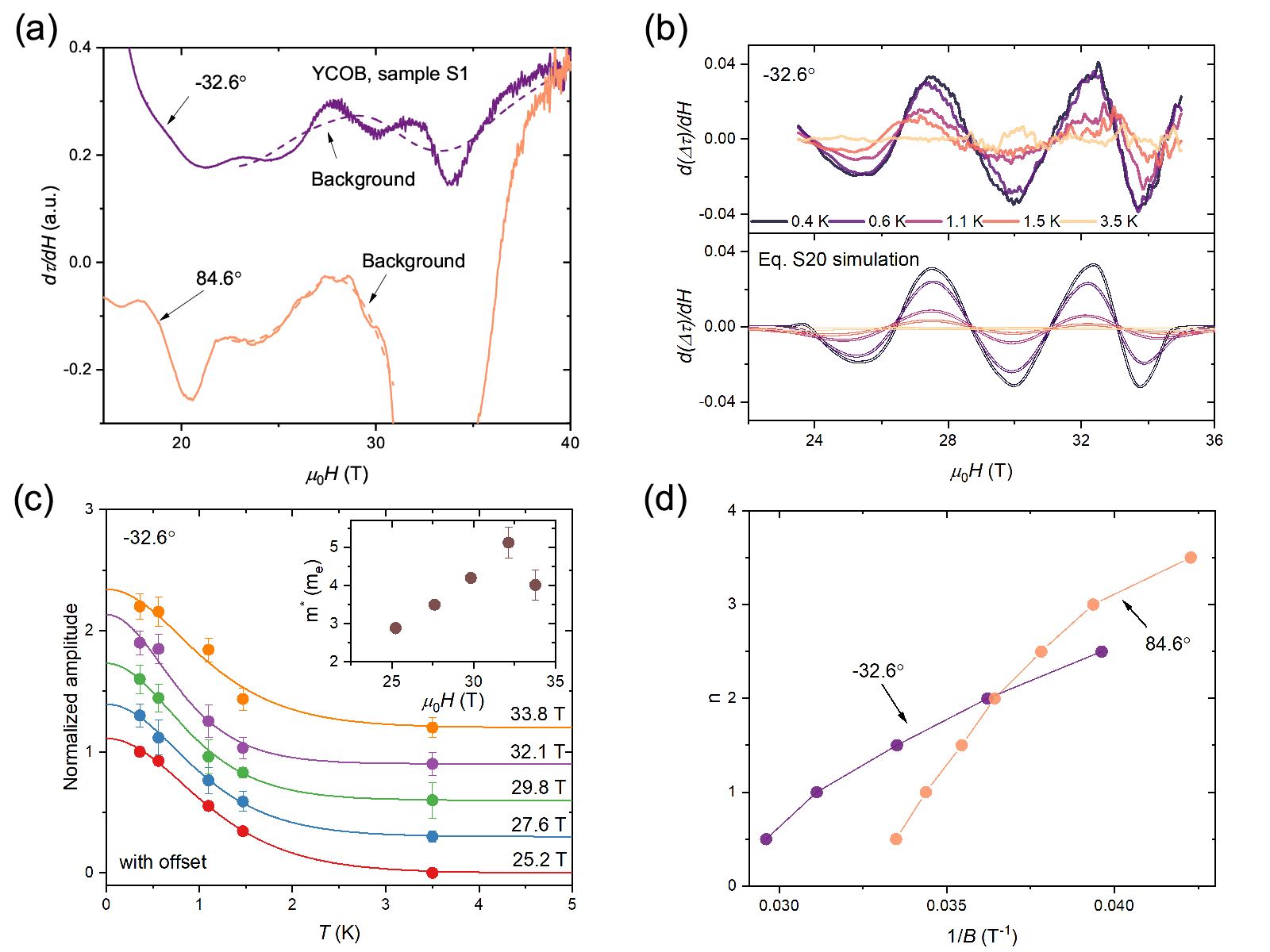}
	
	\caption{{\bf Detailed analysis of torque measurements in YCOB S1.}
		(\textbf{a}) Torque derivative $d\tau/dH$ versus field with dashed curves showing the background for subtraction at $\theta=-32.6^{\circ}$ and $84.6^{\circ}$. The background was obtained by performing Loess smooth on the data, then subtracted to obtain the oscillation component.
		(\textbf{b}) Top panel is the oscillation component isolated from $d\tau/dH$ after background subtraction at $\theta=-32.6^{\circ}$ at different temperatures, and bottom panel is the temperature smeared results by using Eq. \ref{eq_muB_smear} which takes the 0.4 K oscillation data as convolution data and set $g^{\prime}=2 g$.
		(\textbf{c}) The normalized amplitude of oscillation extrema in (b) versus temperature at the fixed field at $\theta=-32.6^{\circ}$. An offset was added to the plots for clarity. The LK fittings are shown in solid lines. Inset: effective mass $m^{*}$ obtained from the LK fitting function. (\textbf{d}) Conventional Landau plot at two angles to check the oscillations are not periodic in $1/B$.
	}
	
	\label{FigS_toranalysis}
\end{figure}

\begin{figure}[!htb]
	\centering
	\includegraphics[width=1\columnwidth]{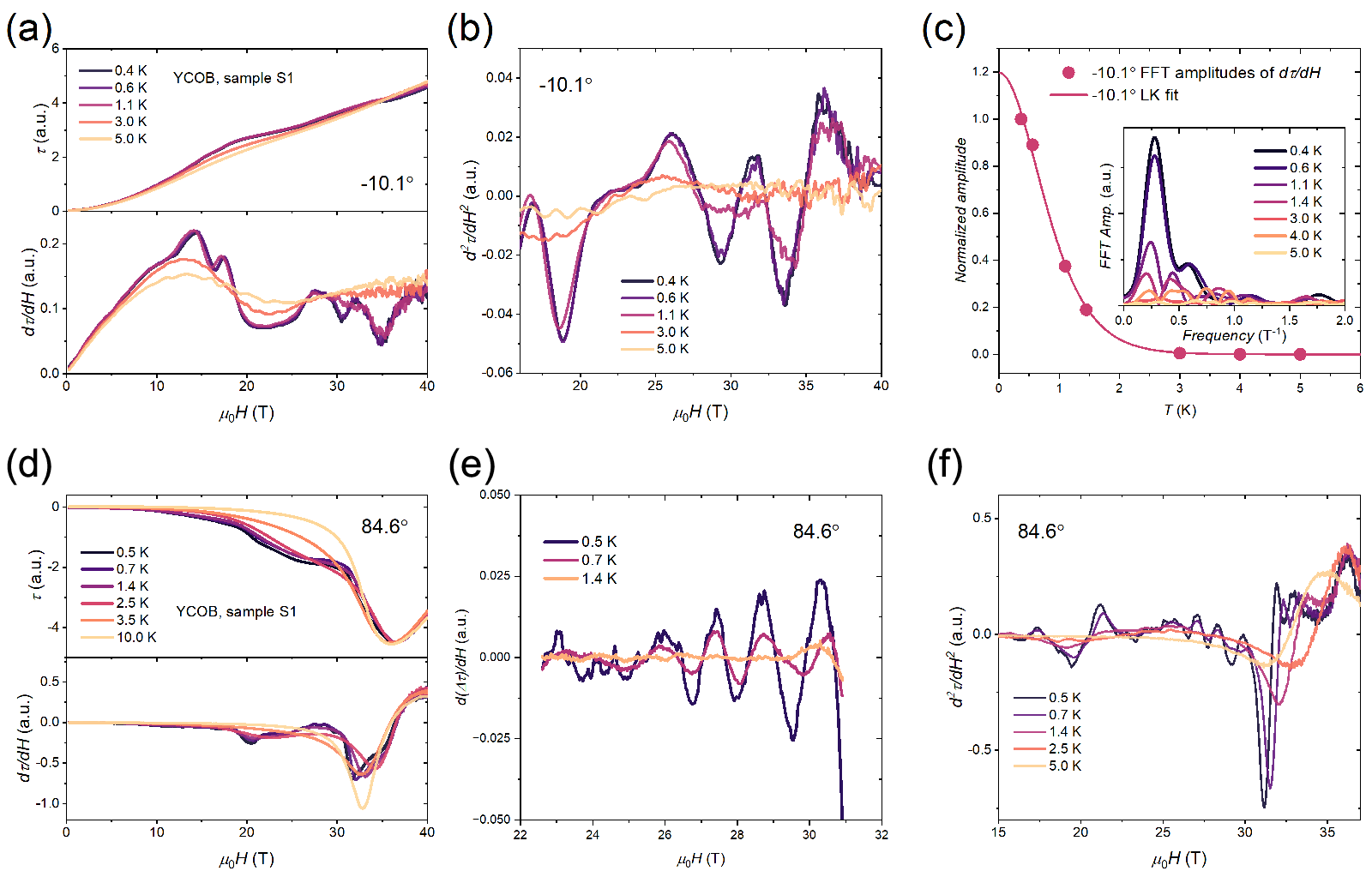}
	
	\caption{{\bf Temperature dependence of -10.1$^{\circ}$ and 84.6$^{\circ}$ data in YCOB S1.}
		(\textbf{a}) (\textbf{d}) Top panel: the measured torque data up to 40 T at different $T$. Bottom panel: the corresponding $d\tau/dH$ plots. (\textbf{b}) The corresponding $d^2\tau/dH^2$ plot at -10.1$^{\circ}$.  (\textbf{c}) LK fitting of normalized FFT amplitudes of $d\tau/dH$ oscillations. The resulting average effective mass for -10.1$^{\circ}$ is $5.2 \ m_e$, which is similar to the value 4$\ m_e$ found in Fig. 3(c) in the main text at -32.6$^{\circ}$. 
		Inset: FFT analysis (See section \ref{sec_FFT}). The FFT window in (\textbf{c}) is [22.4 T, 36.9 T]. (\textbf{e}) The magnetic oscillations at 84.6$^{\circ}$ were obtained from the bottom panel of (\textbf{d}) after subtracting a Loess-smoothed background. (\textbf{f}) The corresponding $d^2\tau/dH^2$ plot at 84.6$^{\circ}$. The sharp dip in (\textbf{f}) around 31 T 
		and the structure above this field shifts with temperature. Its origin is not understood, and it has been excluded from our analyses.
	}
	
	\label{FigS_twoTdepend}
\end{figure}

\begin{figure}[!htb]
	\centering
	\includegraphics[width=\columnwidth]{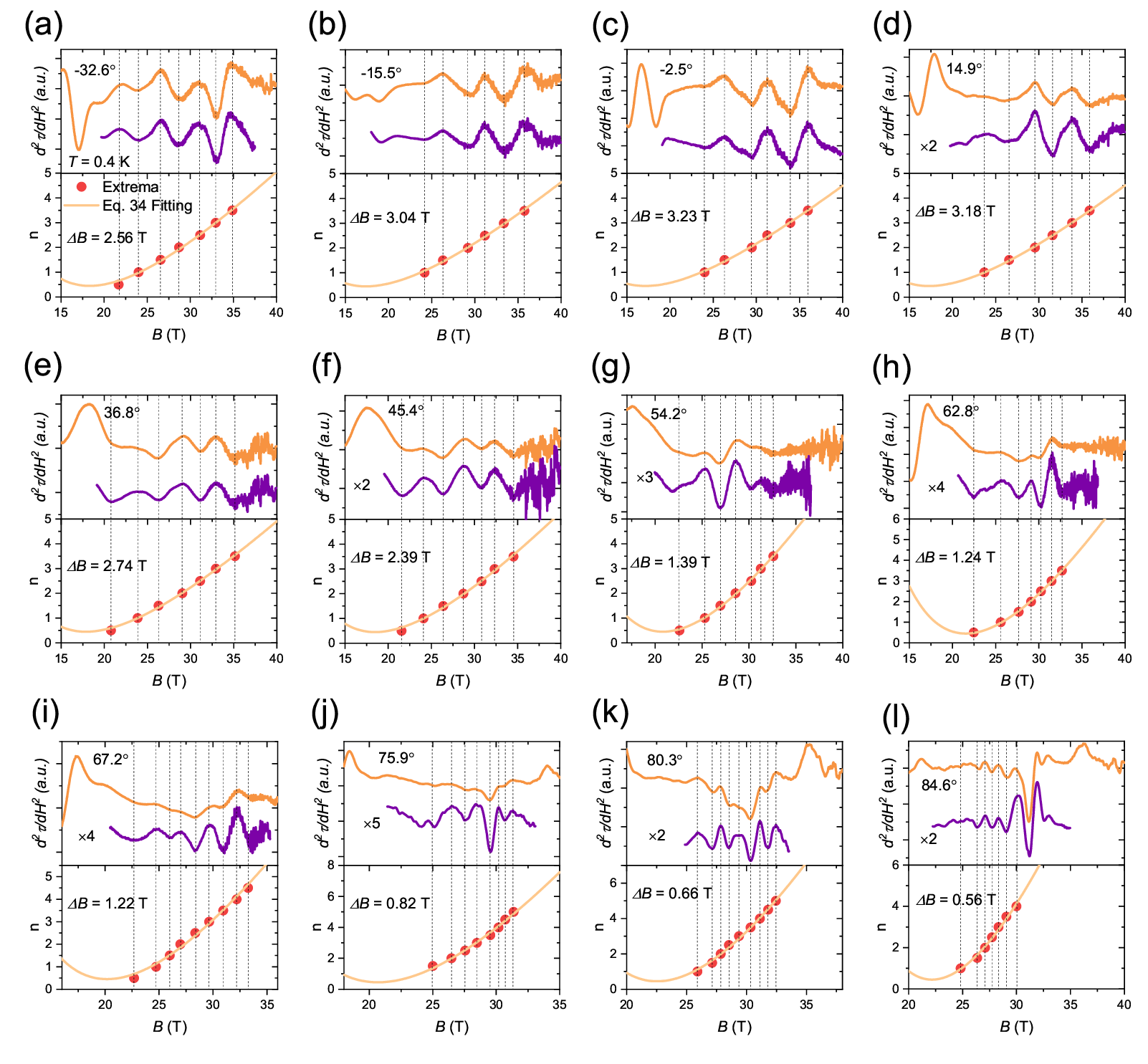}
	
	\caption{{\bf Angular dependence of Landau level plots in YCOB S1. } Each panel is for one tilt angle. Top half of each panel: Orange curves are $d^2\tau/dH^2$ measured at 0.4 K at different angles, and purple curves are the corresponding oscillations after subtracting a Loess smoothed background. ``$\times 3$'' means the oscillation amplitudes are increased by 3 for clarity. Bottom half of each panel: Landau index plots of the $d^2\tau/dH^2$ extrema. Fitting results of $n=(B-B_0)^2/(\Delta B\cdot B)-\varphi^\prime/2\pi$ following Eq. \ref{eq_Landauindex}. $\varphi^\prime/2\pi=-0.45$ is fixed for all angels. The period $\Delta B$ and critical field $B_0$ are determined from fitting.
	}
	
	\label{FigS_Landau}
\end{figure}

\begin{figure}[!htb]
	\centering
	\includegraphics[width=1\columnwidth]{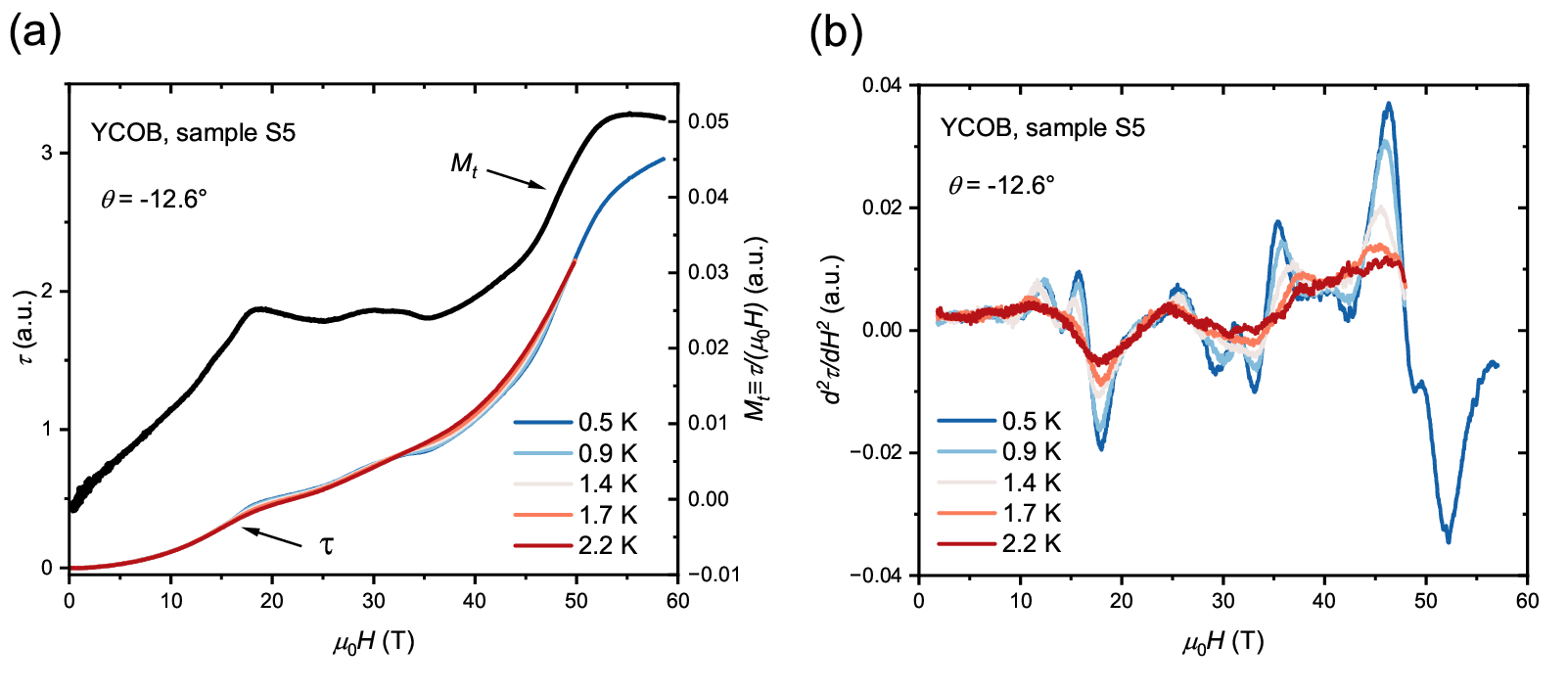}
	
	\caption{{\bf Torque measurement of YCOB S5 in pulsed field. }
		(\textbf{a}) The temperature dependence of YCOB S5 torque signals measured in pulsed magnetic field up to 59 T at -12.6$^{\circ}$. The black curve is the effective magnetization at 0.5 K defined as $\tau/(\mu_0H)$, and the $\frac{1}{9}$ magnetization plateau around 18 T and the following magnetic oscillations show up clearly. (\textbf{b}) The corresponding second derivative of torque signals. The striking features are the abundant magnetic oscillations between the $\frac{1}{9}$ and $\frac{1}{3}$ magnetization plateaus, and the amplitude of oscillations are smeared out by increasing temperature. 
	}
	
	\label{FigS_pulsetor}
\end{figure}

\begin{figure}[!htb]
	\centering
	\includegraphics[width=0.6\columnwidth]{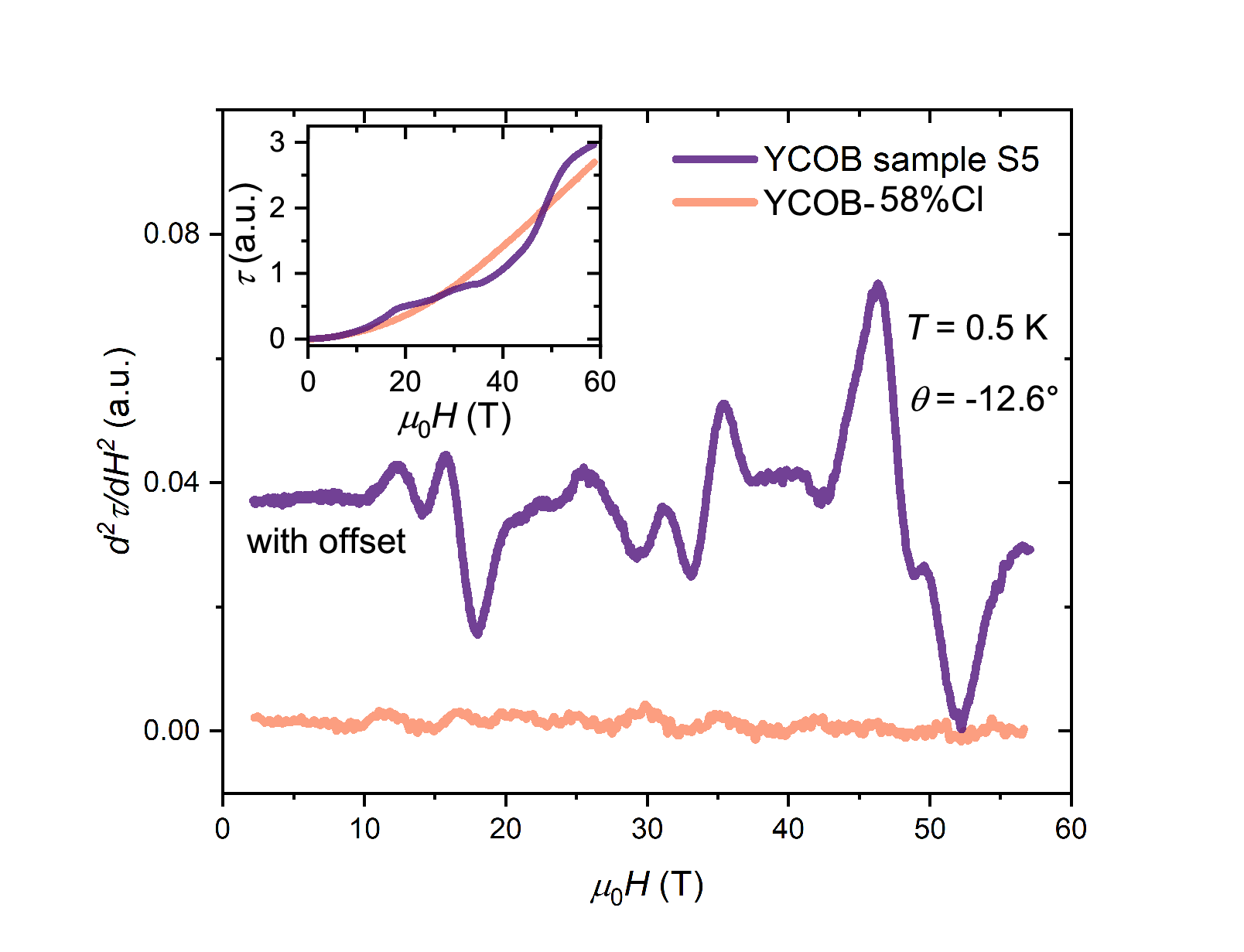}
	
	\caption{{\bf Control experiment.}
		The second derivative signals of YCOB sample S5 and YCOB-58\%Cl were measured simultaneously in pulsed magnetic fields at 0.5 K. The inset shows the raw torque data of both samples. The YCOB S5 signal shows clear plateaus and oscillations, while the data of the control sample (YCOB-58\%Cl) is featureless except for noise. 
	}
	
	\label{FigS_Cl0.5compare}
\end{figure}

\begin{figure}[!htb]
	\centering
	\includegraphics[width=0.6\columnwidth]{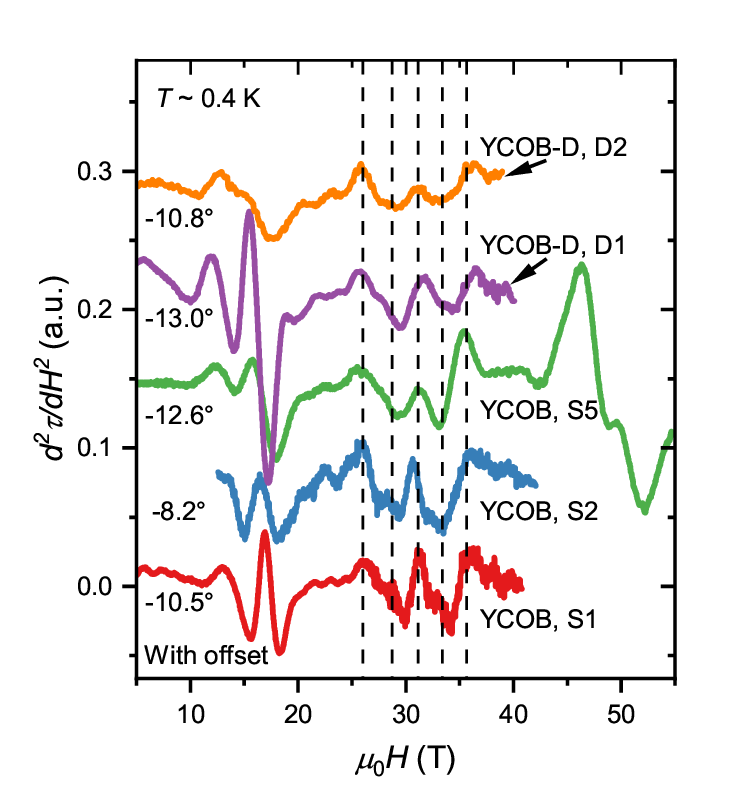}
	
	\caption{{\bf The second derivatives of the magnetic torque in different samples.}
		The susceptibility calculated from $d^2\tau/dH^2$ of five different samples around -10$^{\circ}$ and 0.4 K is shown here. The growth and measurement conditions are shown in \textcolor{blue}{section Materials and Methods} in the main text. The black dash lines marked the extrema of oscillation patterns. Five different samples have similar magnetic oscillations above 20 T, and their extrema can align well with each other. The slight mismatch should come from the difference in angle. The patterns below the $\frac{1}{9}$ magnetization plateau ($\sim$ 18~T) are different from sample to sample, which indicates they may have an extrinsic origin. The variation between samples is discussed in section \ref{sec_diffsample}.
	}
	
	\label{FigS_compare}
\end{figure}

\begin{figure}[!htb]
	\centering
	\includegraphics[width=1\columnwidth]{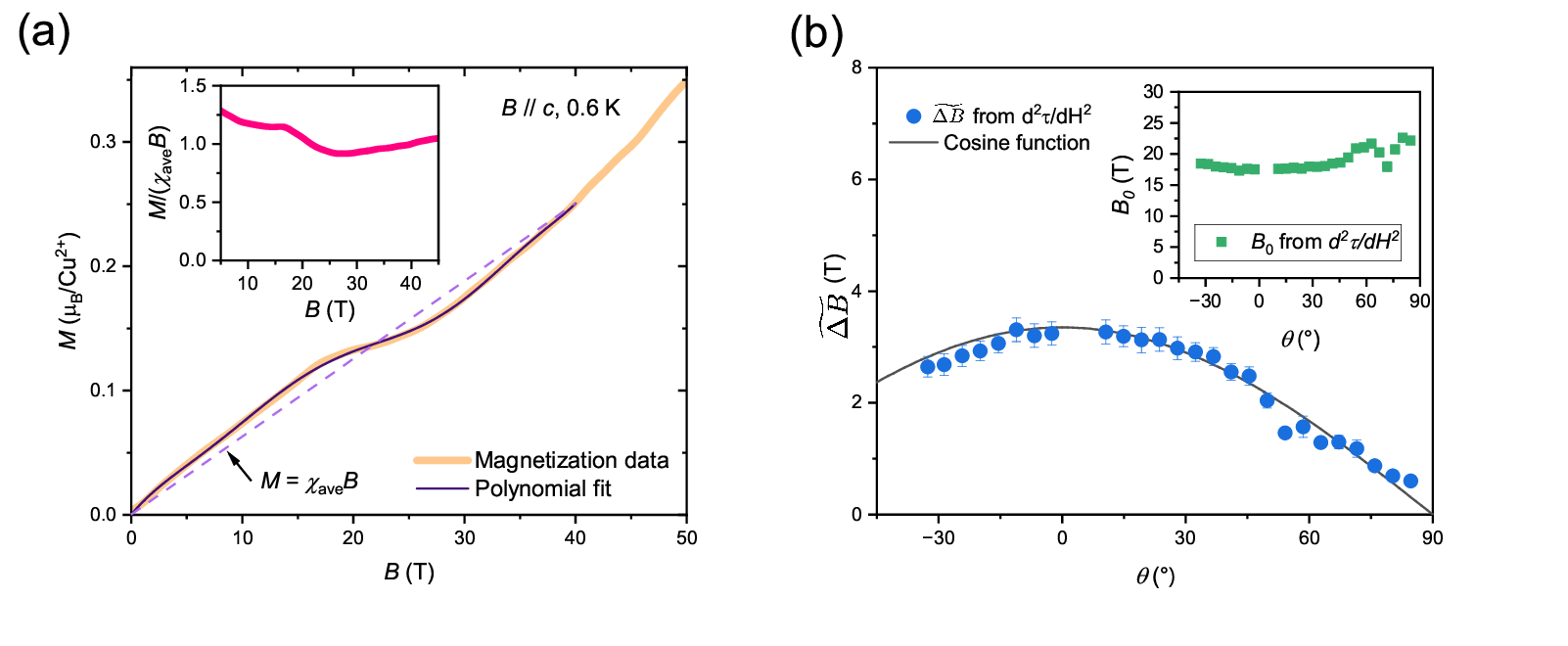}
	
	\caption{{\bf DM Mechanism model fit.}
		(\textbf{a}) The magnetization data (yellow curve) of YCOB sample M1 taken from Fig. 1(b) in the main text. The purple curve is the 8th-order polynomial fit to the magnetization data from 0 to 45 T. The blue dash line represents the function $M=\chi_{ave}B$, with  $\chi_{ave}$ a constant. Inset shows $R(B)=M/(\chi_{ave}B)$ versus magnetic field. 
		Note that its deviation from unity is small, especially between 20 to 40~T, which is the magnetic field range of interest. (\textbf{b}) The angular dependence of parameters $\widetilde{\Delta B}$ and $B_0$. $\widetilde{\Delta B}$ and $B_0$ were obtained from fitting of Landau index plots of $d^2\tau/dH^2$ using Eq.\ref{eq_DMfit}. Theoretically, $\widetilde{\Delta B}$ is predicted to be proportional to $\cos(\theta)$. The black curve is the cosine function that fits the data points. For details, see section \ref{sec_DeltaB}.
	}
	
	\label{FigS_DMmodelfit}
\end{figure}

\end{document}